\documentclass[aps,nofootinbib,superscriptaddress, showpacs,preprintnumbers,  nofootinbibt,twocolumn]{revtex4-2}
\usepackage{eurosym}
\usepackage{amsmath}
\usepackage{bm}
\usepackage{amsfonts}
\usepackage{amssymb}
\usepackage{graphicx}

\def\be{\begin{equation}}
\def\ee{\end{equation}}
\def\bea{\begin{eqnarray}}
\def\eea{\end{eqnarray}}

\begin{document}

\title{Dissipative quintessence, and its cosmological implications}
\author{Tiberiu Harko}
\email{tiberiu.harko@aira.astro.ro}
\affiliation{Department of Physics, Babes-Bolyai University, Kogalniceanu Street, 	Cluj-Napoca 400084, Romania,}
\affiliation{Department of Theoretical Physics, National Institute of Physics and Nuclear Engineering (IFIN-HH), Bucharest, 077125 Romania,}
\affiliation{Astronomical Observatory, 19 Ciresilor Street, Cluj-Napoca 400487, Romania}
\date{\today }

\begin{abstract}
 We consider a generalization of the quintessence type scalar field cosmological models,
by adding a multiplicative dissipative term in the scalar field Lagrangian, which generally is represented in an exponential form. The generalized dissipative Klein-Gordon equation is obtained in a general covariant form in Riemann geometry, from the variational principle with the help of the Euler-Lagrange equations. The energy-momentum tensor of the dissipative scalar field is also obtained from the dissipative Lagrangian, and its properties are discussed in detail. Several applications of the general formalism are presented for the case of the cosmological Friedmann-Lemaitre-Robertson-Walker metric. The generalized Friedmann equations in the presence of the dissipative scalar field are obtained for a specific form of dissipation, with the dissipation exponent represented as the time integral of the product of the Hubble function, and of a function describing the dissipative properties of the scalar field. For this case the Friedmann equations reduce to a system of differential-integral equations, which, by means of some appropriate transformation, can be represented in the redshift space as a first order dynamical system. Several cosmological models, corresponding to different choices of the dissipation function, and of the scalar field potential, are considered in detail. For the different values of the model parameters the evolution of the cosmological parameters (scale factor, Hubble function, deceleration parameter, the effective density and pressure of the scalar field, and the parameter of the dark energy equation of state, respectively), are considered in detail, by using both analytical and numerical techniques. A comparison with the observational data for the Hubble function, and with the predictions of the standard $\Lambda$CDM paradigm is presented for each dissipative scalar field model.  In the large time limit the model describes an accelerating Universe, with the effective negative pressure induced by the dissipative effects associated to the scalar field. Accelerated expansion in the absence of the scalar field potential is also possible, with the kinetic term dominating the expansionary evolution. The dissipative scalar field models describe well the data, with the model free parameters obtained by a trial and error method. The obtained results show the dissipative scalar field model offers an effective dynamical possibility for replacing the cosmological constant, and for explaining the recent cosmological observational data.
\end{abstract}

\pacs{95.36.+x, 98.80.Es, 04.20.Cv, 95.35.+d}
\maketitle

%\tableofcontents

\section{Introduction}

The theory of General Relativity \cite{Ein1,Ein2} is extremely successful in explaining the gravitational phenomena at the level of the Solar System.  A large number of observational and even experimental tests, including the high precision studies of the deflection of light, of the perihelion precession of the planet Mercury, of the Shapiro time delay effect, of the frame-dragging effect, and of the Nordtvedt effect in lunar motion, respectively, have confirmed the validity, and the scientific soundness of the theory  \cite{Will}. Recently, another of the theoretical predictions of General Relativity, was brilliantly confirmed by the experimental detection of the gravitational waves \cite{Grav1}. The gravitational wave studies open a new window into the Universe, leading, to a new perspective on the properties of the black holes, and on the mass distribution of the massive compact astrophysical objects, like, for example, the neutron stars \cite{Grav2}. Very recently, the Event Horizon Telescope (EHT) was able to detect the shadow of the black hole in the center of the M87* galaxy \cite{Ev1,Ev2}, with the observations confirming the general relativistic black hole model. The shadow of a black hole is an important testing ground for the predictions of general relativity,  and of the modified theories of gravity.

However, the improvement of the observational techniques, and the extension of the observations on a much wider scale, led to the unexpected result that for gravitational systems much bigger than the Solar System, General Relativity may not be able to provide an accurate description of their gravitational properties. This situation already appears at the galactic scale, and it becomes even more severe at cosmological scales. Hence, it seems that the theory of General Relativity must face a number of very serious challenges, whose solutions may require a fundamental change in our view of gravity, and of the physical properties of the large scale structures in the Universe.

One of the major discoveries of the past few decades was related to the strong observational evidence indicating that presently the Universe is in a state of accelerating expansion \cite{Ri98,Ri98-1,Ri98-2,Ri98-3, Ri98-4,Hi,Ri98-5, Ri98-6}. These results were obtained from the astrophysical  observations of the distant type Ia supernovae, whose spatial distribution extends up to a redshift of $z\approx 2$. Surprising results also came from the high precision determinations of the temperature fluctuations of the Cosmic Microwave Background Radiation (CMBR), obtained by the Planck satellite \cite{1g,1h}. Moreover, the stunning finding that the matter content of the Universe consists of only 5\% baryonic matter has also been decisively confirmed by multiple observations. Hence, the present day observational situation in cosmology convincingly indicates that 95\% of the total composition of the Universe resides in the form of two main (and mysterious) constituents, dark energy and dark matter, respectively. On the other hand, it is important to point out that a cosmic fluid, formed of normal matter, and obeying a perfect fluid type equation of state, cannot trigger and sustain the accelerated expansion of the Universe \cite{Sen}.

A theoretical interpretation of the cosmological observations can be achieved through the  reintroduction in the Einstein  field equations of the cosmological constant $\Lambda$, first proposed by Einstein in 1917 \cite{Einb}, in order to obtain a static cosmological model. For the interesting history of the cosmological constant, of its rejections and returns, as well as of its many possible interpretations see \cite{W1b,W2b,W3b,W4b,W5b}. The cosmological model, obtained by adding to the Einstein field equations the cosmological constant $\Lambda$, as well as a cold dark matter component, is called the $\Lambda$CDM model. Presently, the $\Lambda$CDM model represents one of the main theoretical instruments used for the comprehension of the cosmic dynamics, and for the interpretation of the observational data.

The $\Lambda$CDM paradigm gives very good fits to the observations. But it lacks a convincing theoretical foundation, which is first of all related to the interpretational  problems related to the cosmological constant itself. This makes the physical basis of the $\Lambda$CDM model problematic. Moreover, the $\Lambda$CDM model is recently facing another major problem. Measurements of the Hubble constant in the early Universe indicate a value of the order of $H_0 < 69$ km/ s /Mpc, while local
measurements give  $H_0 > 71$ km/ s/ Mpc \cite{Anc}. The contradictions between the values obtained in the measurements of the Hubble constant are known as the
Hubble tensions, and their extents depend on the used data sets.

Therefore, to obtain theoretically consistent description of the Universe, several approaches have been proposed, which try to solve the cosmological constant problem by assuming some alternative explanations of the cosmic dynamics, which could be described as the dark component, the dark gravity, and the dark coupling models \cite{LoHa}.

One of the important alternatives to the $\Lambda$CDm models are represented  by the dark components model \cite{Rev1,Rev2,Rev3,Rev4, Rev5}. In the framework of this approach one postulates that the basic constituents of the Universe are the dark energy, and the dark matter, respectively, whose physical properties could explain, at least at a phenomenological level,  the cosmological observations. Many proposals for the physical nature of these two dark constituents have been considered, and investigated in detail. Perhaps the simplest dark energy model can be obtained by using the quintessence type theories \cite{quint1b,quint2b,quint3b,quint4b, quint5b}. In the quintessence theory the cosmological evolution of the Universe is fully determined, and described, by a single scalar field $\phi$, in the presence of a self-interaction potential $V(\phi)$.  The simplest gravitational action for the quintessence models is given by,
\be
S=\int{\left[\frac{M_{p}^2}{2}R-\partial_{\mu} \phi\partial ^\mu-V(\phi)\right]\sqrt{-g}d^4x},
\ee
where by $R $ we have denoted the Ricci scalar, while $M_p$ represents the Planck mass. The cosmological energy density and pressure of the quintessence scalar field are obtained as $\rho_Q=\dot{\phi}^2/2+V(\phi)$, and $p_Q=\dot{\phi}^2/2-V(\phi)$ \cite{Cop}, giving for the equation of state $w$ of the quintessence field the expression $w=p_Q/\rho_Q=\left(\dot{\phi}^2/2-V(\phi)\right)/\left(\dot{\phi}^2/2+V(\phi)\right)$. Quintessence type cosmological models have been very successful in interpreting, and explaining important characteristics of the cosmic evolution. For recent reviews on quintessence theories se \cite{RevQ1,RevQ2}. In particular, quintessential cosmological models can solve the $\sigma_8$ tension, by allowing  the conformal coupling of a single dark energy scalar field to dark matter through a constant coupling \cite{Sol1}.  The Hubble tension can be alleviated by considering a quintessence field that transitions from a matter-like to a cosmological constant behavior between the recombination and the present time \cite{Sol2}. The discrepancy between the local measurements of $H_0$ and that inferred from the cosmic microwave background observations can be reconciled by assuming the existence of an electroweak axion in the minimal supersymmetric standard model, with the axion energy density identified with the observed dark energy \cite{Sol3}. The best-fit of the dark energy parameters was used to reconstruct the quintessence Lagrangian in \cite{Sol4}. Due to the derived late phantom behavior of  $w(z)$, the reconstructed quintessence models have a negative kinetic term. The possibility of alleviating both the $H_0$ and the $\sigma_8$ tensions simultaneously by means of the Albrecht-Skordis ``quintessence'' potential was considered in \cite{Sol5}. The quintessence field can reduce the size of the sound horizon  $r_s$,
while suppressing the power in matter density fluctuations before it dominates the present day energy density. For some recent works on the cosmological implications of the quintessence model see \cite{Sol6a, Sol7a, Sol8a, Sol9a, Sol10a, Sol11a, Sol12a, Sol13a}.

In \cite{Am} the coupled quintessence (CQ) model was proposed, in which the scalar field $\phi$ and the dark matter fluid interact with each other through a source term $Q_\nu$, which appears in the conservation equations as
\be
\nabla _{\mu}T^{\mu}_{\nu (\phi)}=-Q_{\nu}, \nabla _{\mu}T^{\mu}_{\nu (m)}=Q_{\nu},
\ee
where $T^{\mu}_{\nu (\phi)}$ and $T^{\mu}_{\nu (m)}$ are the energy-momentum tensors of the scalar field, and of the dark matter, respectively. It was also suggested \cite{Am} that the source term can be given by $Q_\nu=-\kappa \beta (\phi)T_{(m)}\nabla _\nu \phi$, where $T_{(m)}$ is the trace of the matter energy-momentum tensor, and $\beta(\phi)$ is the coupling function that determines the strength of the interaction.

Several other scalar field models have been explored from a cosmological perspective. In a class of string theories, depending on the form of the tachyon potential, the tachyon scalar field can act as a source of the  dark energy \cite{Pad1,Pad2, Sen1, Sen2}. The effective Lagrangian for the tachyon scalar field is given by,
\be
L=-V(\phi)\sqrt{1+\partial _\mu \phi \partial ^\mu \phi},
\ee
where $\phi$ is the tachyon scalar field, and $V(\phi)$ is its potential. The energy density and the pressure of the tachyon field are given by,
\be
\rho_T=\frac{V(\phi)}{\sqrt{1-\dot{\phi}^2}}, p_T=-V(\phi)\sqrt{1-\dot{\phi}^2},
\ee
giving for the parameter of the equation of state $w_T$ the expression,
\be
w_T=\frac{p_T}{\rho_T}=\dot{\phi}^2-1.
\ee
For recent studies on the tachyonic field cosmology see \cite{T1, T2,T3}.

Another interesting scalar field theory that was introduced to explain the cosmological observations is the k-essence scalar field model of dark energy. The main characteristic of the model is the presence of a scalar
field with a non-canonical kinetic energy term. The scalar field action for k-essence is a
function of the field $\phi$, and of $\chi = \dot{\phi}^ 2/2$,  and it is given by \cite{K0, K01, K02},
\be
S=\int{p_{DE}\left(\phi,\chi\right)\sqrt{-g}d^4x},
\ee
where the Lagrangian density corresponds to the pressure of a scalar field with a non-canonical kinetic term, given by,
\be
p_K=f(\phi)\left(-\chi+\chi^2\right),
\ee
The energy-density of the k-essence field is given by,
\be
\rho_K=f(\phi)\left(-\chi +3\chi^2\right).
\ee
For the parameter $w_K$ of the equation of state of the k-essence field we obtain,
\be
w_K=\frac{\chi-1}{3\chi -1}.
\ee
For the cosmological applications, and implications, of the k-essence models, see \cite{K1, K2,K3,K4,K5}, and references therein.

Finally we would like to mention the dilaton scalar field model, which is an attempt to solve the dark energy problem by using string theory \cite{D1, D2, D3}. For the dilation scalar the energy density and the pressure are given by,
\be
\rho_D=-\chi+3ce^{\lambda \phi}\chi ^2,p_D=-\chi+ce^{\lambda \phi}\chi ^2,
\ee
where $c$ and $\lambda$ are constants. The parameter of the equation of state of the dilaton scalar field is given by,
\be
w_D=\frac{1-ce^{\lambda \phi}}{1-3ce^{\lambda \phi}}.
\ee

There are also some other approaches to the cosmological phenomenology. For example, in the dark gravity approach it is assumed the gravitational interaction itself is modified on the galactic and cosmological scales. One possibility to modify gravity is to go beyond the Riemannian geometry of general relativity,  and to use more general geometries to describe gravity. In this direction theories in the presence of torsion \cite{Tor1,Tor2,Tor3,Tor4}, of nonmetricity  \cite{Nest1,Nest2,Nest3, Ghil1,Ghil2,Ghil3}, or in the Weitzenb\"{o}ck geometry \cite{We1,We2} have been intensively investigated. The third theoretical avenue for explaining the cosmological phenomenology is the dark coupling approach, which assumes that ordinary matter can couple with geometry, through a curvature - matter coupling.  The existence of such a coupling could explain the accelerated expansion of the Universe, as well as the dark matter problem  \cite{Co1,Co2,Co3,Co4,Co5}. For reviews of the modified gravity theories see \cite{R1,R2,R3,R4,R5, e8}.

Decay processes plays a central role in a wide range of phenomena, including
nuclear fission or optical emission. Dissipation also appears in quantum
systems, and it is a consequence of the dissipative interaction of the
quantum system with its environment \cite{1}. Energy decay is usually
considered as a consequence of a thermodynamic system exchanging energy
irreversibly with its environment, usually assumed to be a thermal bath.
However, there are energy decay processes that cannot be explained by
assuming a direct coupling to a thermal bath \cite{2}.

A class of dissipative effects, the bulk and shear viscous processes, have been extensively investigated in astrophysical and cosmological settings, and they are assumed
to play an important role in the early evolution of the Universe.  A cosmic fluid with bulk viscous pressure, in the presence of  quintessence field can trigger the accelerated expansion phase of the Universe \cite{Chim}. The presence of bulk viscosity could also solve the coincidence problem of cosmology. The bulk viscous Chaplygin gas model was considered in \cite{Chap}.  A recent investigation of a unified cosmic fluid scenario in the presence of bulk viscosity, with the coefficient of the bulk viscosity having a power law evolution was carried out in \cite{Bulk}. Considering such a general bulk viscous scenario, the observational constraints using the latest cosmological datasets were obtained, and their behaviour was analyzed at the level of both background solutions, and cosmological perturbations. The observational analyses did show that a non-zero bulk viscous coefficient is always favored. Moreover,  some of the bulk viscous models can weaken the current $H_0$ tension for some datasets. But from the Bayesian evidence analysis, it follows that the $\Lambda$CDM model is favored over the cosmological models with bulk viscosity.

A problem less investigated in the physical literature is the possibility of
a Lagrangian description of dissipative phenomena. In this respect one must
make a clear distinction between physical (standard) and mathematical
(non-standard) Lagrangians. A physical Lagrangian is a Lagrangian function
that can be represented as the difference between a kinetic energy term, and
a potential energy term. Other Lagrangians, which also give the correct
equation of motion, but which cannot be represented as the difference of a
kinetic and potential term, are called mathematical, or non-standard,
Lagrangians. For example, the equation of motion of the damped oscillations,
describing the motion of a single particle of mass $m$ in an external field
with potential $V(x)$ and in the presence of friction, can be obtained, via
the Euler-Lagrange equations,
\begin{equation}
\frac{\partial L}{\partial x}-\frac{d}{dt}\frac{\partial L}{\partial \dot{x}}%
=0,
\end{equation}%
from the physical Lagrangian \cite{Cis, Ber}
\begin{equation}
L=e^{\gamma t}\left( \frac{1}{2}m\dot{x}^{2}-V(x)\right) ,
\end{equation}%
where a dot denotes the derivative with respect to the time $t$, and is
given by
\begin{equation}
\ddot{x}+\gamma \dot{x}+\frac{1}{m}\frac{dV(x)}{dx}=0.
\end{equation}%
One can also construct a Hamiltonian for the damped oscillator in the
standard way. By defining $p=\partial {L}\partial {\dot{x}=me}^{\gamma t}%
\dot{x}$, $\mathcal{H}=p\dot{x}-L$, one immediately obtains
\begin{equation}
\mathcal{H}=e^{-\gamma t}\frac{p^{2}}{2m}+e^{\gamma t}V(x),\mathcal{H}%
=e^{\gamma t}\left( \frac{1}{2}m\dot{x}^{2}+V(x)\right) .
\end{equation}

It is important to note that $\mathcal{H}$ as defined above is not the
energy of the system, which is still defined as $E=m\dot{x}^{2}/2+V(x)$, and
satisfies the relation $dE/dt=F_{d}\dot{x}$, where $F_{d}=-m\gamma \dot{x}$
is the dissipative force. Therefore, $\mathcal{H}=e^{\gamma t}E$, and it
cannot be interpreted as the energy of the system \cite{3}. It is also
interesting to point out that the equation $\ddot{x}+k\dot{x}=0$ can be
derived from the physical (standard) Lagrangian $L=e^{kt}\dot{x}^{2}/2$, as
well as from the non-physical Lagrangians $L=1/\left( e^{2kt}\dot{x}%
+e^{kt}\right) $, $L=\dot{x}\ln \left\vert \dot{x}\right\vert -kx$, or $%
L=\left( \dot{x}^{\nu }+e^{-\nu kt}\right) ^{1/\nu }$, respectively \cite%
{Cis}.

From a mathematical point of view, the Klein-Gordon equation describing the
cosmological evolution of a scalar field in a
Friedmann-Lemaitre-Robertson-Walker (FLRW) geometry belongs to the general
class of equations of the form,
\begin{equation}
\ddot{x}+F(t)\dot{x}+g^{\prime }(x)=0,  \label{3e}
\end{equation}%
where $F$ and $g$ are arbitrary functions of time. Eq.~(\ref{3e}) can be
derived from the dissipative physical Lagrangian, =0,%
\begin{equation}
L=e^{\int {F(t)dt}}\left( \frac{1}{2}\dot{x}^{2}-g(x)\right) ,
\end{equation}%
with the use of the Euler-Lagrange equations, by taking into account that $%
\partial L/\partial \dot{x}=\dot{x}e^{\int {F(t)dt}}$, $d\left( \partial
L/\partial \dot{x}\right) /dt=e^{\int {F(t)dt}}\left[ \ddot{x}+F(t)\dot{x}%
\right] $, and $\partial L/\partial x=-g^{\prime }(x)e^{\int {F(t)dt}}$,
respectively.

In the Minkowskian space a natural dissipative extension of the scalar
field, and of the Klein-Gordon equation can be considered by adopting for
the Lagrangian density the expression \cite{Scom},
\begin{equation}
L_{\phi }=e^{k_{\alpha }x^{\alpha }}\left( \frac{1}{2}\partial _{\mu }\phi
)\partial ^{\mu }\phi -V(\phi )\right) ,
\end{equation}%
where $k_{\alpha }$ are constants. From the Euler-Lagrange equations, $%
\partial L/\partial \phi -\partial _{\mu }\left( \partial L/\partial \phi
_{,\mu }\right) =0$, where $\phi _{,\mu }=\partial \phi /\partial x^{\mu }$,
we obtain the equation of motion of the dissipative scalar field as,
\begin{equation}
\partial _{\mu }\partial ^{\mu }\phi +k_{\mu }\partial ^{\mu }\phi
+V^{\prime }(\phi )=0.
\end{equation}

In the above equation, by analogy with the equations of motion for the
damped oscillators, one could interpret the term $k_{\mu }\partial ^{\mu
}\phi $ as corresponding to a dissipative friction term.

It is the goal of the present paper to extend, and formulate the variational
formulation of the dissipative scalar field by using a fully covariant
approach in the Riemannian geometric framework, and to formulate the
dissipative Klein-Gordon equation for the scalar field in a general
covariant form. The dissipation is introduced in the Lagrangian via a
dissipation exponent $\Gamma $, assumed to be, in general, a function of the
metric tensor, of the scalar field, and of the coordinates of the base
space-time manifold. By using the analogy with the simple damped harmonic
oscillator, the dissipative Lagrangian is obtained then by multiplying the
Lagrangian of the "ideal" scalar field with the exponential of $\Gamma $, so
that the new Lagrangian is constructed as the product of $e^{\Gamma }$, and
the standard Lagrange function of the scalar field. The generalized
Klein-Gordon equations are obtained in a fully covariant form for the case
of the dissipation exponent having various functional forms. Particular
cases of the dissipative Klein-Gordon equation are also discussed in detail.

Once the Lagrangian density $L_{\phi }$ of the scalar field is known, the
basic physical properties of the field can be obtained from the
energy-momentum tensor, which can be straightforwardly obtained from $%
L_{\phi }$ through variation with respect to the metric. We obtain the
general form of the energy-momentum tensor of the dissipative scalar field,
which involves the presence of a new tensor, the dissipation tensor, which
gives a new, significant, and important contribution to both the energy
density and to the pressure of $\phi $.

As a particular example of the general formalism we consider the case in
which the dissipation \ exponent can be expressed as the invariant integral
of the divergence of a four-vector $u^{\lambda }$, and of an arbitrary
function $Q\left( x^{\mu }\right) $, which we call the dissipation function,
having the mathematical representation given by $\Gamma \left( g_{\alpha
\beta },x^{\mu }\right) =\int_{\Omega }\nabla _{\lambda }u^{\lambda }Q\left(
x^{\mu }\right) \sqrt{-g}d^{4}x$. This case is analyzed in detail, and the
dissipative Klein-Gordon equation, as well as the corresponding
energy-momentum tensor, are obtained in a covariant form.

We extensively apply the obtained results to generalize, and extend, the
standard cosmological scalar field models, which have been successfully used
to explain the recent acceleration of the Universe. To do this, we restrict
our analysis to the case of the flat, isotropic and homogeneous
Friedmann-Lemaitre-Robertson-Walker (FLRW) geometries. For the specific
cosmological applications, we assume that the dissipation exponent can be
expressed as the integral of the product of the Hubble function $H(t)$, and
of the dissipation function $Q(t)$, in the form $\Gamma (t)=3\int H(t)Q(t)dt$%
. We obtain the generalized dissipative Klein-Gordon equation, as well as
the corresponding energy-momentum tensor for the field in a cosmological
setting. With the help of these quantities, the generalized Friedmann
equation, describing the cosmological evolution in the presence of the
dissipative scalar field are obtained.

In order to test the cosmological viability of the dissipative scalar field
model, we consider several explicit cosmological models, corresponding to
various choices of the dissipation function $Q$. First, the existence of the
de Sitter type solution for this model is proven. Then, several classes of
cosmological models, corresponding to a constant $Q$, and to a redshift
dependent dissipation, are considered in detail. Models in which the kinetic
term, and the potential term of the field can be neglected are investigated
numerically. In each case a comparison with the observational data for the
Hubble function, and with the standard $\Lambda $CDM model are performed,
and it is shown that the models give a good description of the data. The
obtained results indicate that the dissipative scalar field model can be
considered as a viable extension of the standard quintessence type
cosmological models. This model also offers a firm theoretical foundation,
via its variational principle, to different classes of scalar field
cosmologies, and allows the possibility of their rigorous generalization.

The present paper is organized as follows. In Section~\ref{sect1}, after
briefly reviewing the basic theory of the ideal cosmological scalar fields,
we introduce the dissipative scalar field in the FLRW geometry via the
variational principle. The generalized Klein-Gordon equations are obtained
for a dissipation exponent given by $\Gamma (t)=3\int H(t)Q(\phi (t),t)dt$,
with several particular cases considered. The covariant form of the
Klein-Gordon equation is obtained, and a particular case is investigated in
detail. The Einstein, and the generalized Friedmann equations are obtained
in Section~\ref{sect2}. Simple cosmological applications of the dissipative
scalar field model are investigated in Section~\ref{sect3}, by considering
some simple forms of the scalar field, and by assuming a constant $Q$.
Comparisons with the observational data, and the standard $\Lambda$CDM model
are also performed. A cosmological model with a dynamic, redshift dependent
dissipation function $Q$ is analyzed in Section~\ref{sect4} Finally, we
discuss and conclude our results in Section~\ref{sect5}.

In the present paper we use the Landau-Lifshitz \cite{LaLi} sign and geometric
conventions.

\section{The dissipative Klein-Gordon equation}\label{sect1}

In this Section we will introduce the basic variational formalism for the
description of the dissipative scalar fields. After briefly reviewing the
non-dissipative case, as well as its cosmological applications, we proceed to
the systematic presentation of the various forms of the dissipative
Klein-Gordon equation, and of their cosmological formulations.

\subsection{Non-dissipative (ideal) scalar fields}

In a Riemannian geometry, the action for an ideal scalar field with
self-interaction potential $V\left( \varphi \right) $ is given by

\begin{equation}
S_{\phi }=\int L_{\phi }d^{4}x=\int \left[ \frac{1}{2}g^{\alpha \beta }\frac{%
\partial \phi }{\partial x^{\alpha }}\frac{\partial \phi }{\partial x^{\beta
}}-V\left( \phi \right) \right] \sqrt{-g}d^{4}x,
\end{equation}%
where $g^{\alpha \beta }$ are the components of the metric tensor, and $-g$
is its determinant.

The Euler-Lagrange equations, giving the minimum of the action, are,
\begin{equation}\label{ELin}
\frac{\partial }{\partial x^{\alpha }}\frac{\partial L_{\phi }}{\partial
\phi _{,\alpha }}-\frac{\partial L_{\phi }}{\partial \phi }=0.
\end{equation}

Therefore, we obtain,
\begin{equation}
\frac{\partial }{\partial x^{\alpha }}\left( \sqrt{-g}g^{\alpha \beta }\frac{%
\partial \phi }{\partial x^{\beta }}\right) +\frac{dV\left( \phi \right) }{%
d\phi }\sqrt{-g}=0.
\end{equation}

But, it is easy to check that \cite{LaLi},
\begin{equation}
\frac{1}{\sqrt{-g}}\frac{\partial }{\partial x^{\alpha }}\left( \sqrt{-g}%
g^{\alpha \beta }\frac{\partial \phi }{\partial x^{\beta }}\right) =\nabla
_{\alpha }\nabla ^{\alpha }\phi =\Box \phi ,
\end{equation}%
where by $\nabla _{\alpha }$ we have denoted the covariant derivative with
respect to the metric, and $\square =\nabla _{\alpha }\nabla ^{\alpha }$ is
the d'Alembert operator. Hence, we obtain the covariant Klein-Gordon
equation in Riemann geometry, describing the dynamics of an ideal,
non-dissipative scalar field, as given by,
\begin{equation}
\square \phi +\frac{dV\left( \phi \right) }{d\phi }=0.  \label{1}
\end{equation}

The energy-momentum tensor is defined generally by the relation \cite{LaLi},
\begin{equation}
T_{\alpha \beta }=\frac{2}{\sqrt{-g}}\frac{\delta \left[ \sqrt{-g}L\left(
g_{\alpha \beta },\phi \right) \right] }{\delta g^{\alpha \beta }},
\end{equation}%
where $L\left( g_{\alpha \beta },\phi \right) $ is any physical Lagrangian
function, which is assumed to be independent on the derivatives of the
metric tensor. Hence, for the energy-momentum tensor of the ideal scalar
field we obtain,
\begin{equation}
^{(\phi )}T_{\alpha \beta }=\phi _{,\alpha }\phi _{,\beta }-\left( \frac{1}{2%
}g^{\mu \nu }\phi _{,\mu }\phi _{,\nu }-V(\phi \right) g_{\alpha \beta }.
\end{equation}

With the use of the Klein-Gordon equation (\ref{1}), one could immediately
check that the energy-momentum tensor of the ideal scalar field satisfies
the conservation condition $\nabla _{\beta }\left( ^{(\phi )}T_{\alpha
}^{\beta }\right) =0$.

$T_{\alpha \beta }$ can be recast in the standard form of a perfect fluid,
\begin{equation}
T_{\alpha \beta }=\left( \rho _{\phi }+p_{\phi }\right) U_{\alpha }U_{\beta
}-p_{\phi }g_{\alpha \beta },
\end{equation}%
where we have introduced the energy density $\rho _{\phi }$ and the pressure
$p_{\phi }$ of the scalar field, defined as,
\begin{equation}
\rho _{\phi }=\frac{1}{2}g^{\mu \nu }\phi _{,\mu }\phi _{,\nu }+V(\phi
),p_{\phi }=\frac{1}{2}g^{\mu \nu }\phi _{,\mu }\phi _{,\nu }-V(\phi ),
\end{equation}%
and the effective four-velocity of the field $U_{\alpha }$, given by,
\begin{equation}
U_{\alpha }=\frac{\phi _{,\alpha }}{\sqrt{g^{\mu \nu }\phi _{,\mu }\phi
_{,\nu }}},
\end{equation}%
respectively.

\subsubsection{Application: the case of the FLRW geometry}

The standard flat, isotropic and homogeneous cosmological FLRW metric is
given by
\begin{equation}
ds^{2}=c^{2}dt^{2}-a^{2}(t)\left( dx^{2}+dy^{2}+dz^{2}\right) ,  \label{FLRW}
\end{equation}%
where $a(t)$ is the scale factor. Then, we have $\sqrt{-g}=a^{3}(t)$.
Furthermore, we assume $\phi =\phi \left( t\right) $. An important
observational quantity, the Hubble parameter, is defined as $H(t)=\dot{a}%
(t)/a(t)$.

The Lagrangian of the time-dependent ideal scalar field is given by,
\begin{equation}
L_{\phi }=a^{3}\left( \frac{1}{2}\dot{\phi}^{2}-V(\phi )\right) .
\end{equation}

Thus, Eq. (\ref{ELin}) gives,
\begin{equation}
\frac{1}{a^{3}}\frac{d}{dt}\left( a^{3}\frac{d\phi }{dt}\right) +\frac{%
dV(\phi )}{d\phi }=0,
\end{equation}%
or, equivalently,
\begin{equation}
\ddot{\phi}+3H\dot{\phi}+V^{\prime }(\phi )=0.
\end{equation}%
The energy density and the pressure of the cosmological scalar field become,
\begin{equation}
\rho _{\phi }=\frac{1}{2}\dot{\phi}^{2}+V(\phi ),p_{\phi }=\frac{1}{2}\dot{%
\phi}^{2}-V(\phi ).
\end{equation}

\subsection{The dissipative scalar field}

We consider now the variational formulation of the dissipative Klein-Gordon
equation. We begin our analysis with the simple case of the cosmological
scalar fields, which are further generalized to a full covariant formalism.

\subsubsection{Dissipation in the FLRW geometry}

The Lagrangian of a dissipative scalar field in a FLRW type geometry can be
taken as,
\begin{equation}
L_{\phi }=a^{3}e^{3\int H(t)Q(t)dt}\left( \frac{1}{2}\dot{\phi}^{2}-V(\phi
)\right) .  \label{L1}
\end{equation}

In the following we will call the function $\Gamma (t)=3\int {H(t)Q(t)dt}$
\textit{the dissipation exponent}, where $Q=Q(t)$ is \textit{the dissipation
function}. Then, the Euler-Lagrange equation giving the minimum of the
action constructed with the help of the Lagrangian (\ref{L1}) takes the
form,
%\begin{equation}
%\frac{d}{dt}\left( a^{3}e^{3\int H(t)Q(t)dt}\frac{d\phi }{dt}\right) +\frac{%
%dV\left( \phi \right) }{d\phi }a^{3}e^{3\int H(t)Q(t)dt}=0,
%\end{equation}%
%or,
\begin{eqnarray}
&&\Bigg[3a^{2}\dot{a}\frac{d\phi }{dt}+3a^{3}H(t)Q(t)\frac{d\phi }{dt}+a^{3}%
\frac{d^{2}\phi }{dt^{2}}+\frac{dV\left( \phi \right) }{d\phi }a^{3}\Bigg]
\notag \\
&&\times e^{3\int H(t)Q(t)dt}=0,
\end{eqnarray}%
giving, for $a\neq 0$,
\begin{equation}\label{DisKG}
\ddot{\phi}(t)+3H(t)\left( 1+Q(t)\right) \dot{\phi}(t)+\frac{dV\left( \phi
\right) }{d\phi }=0.
\end{equation}%
%where we have used the relation $\frac{d}{dt}\int H(t)Q(t)dt=$ $H(t)Q(t)$.

Hence, the function $Q(t)$ acts as a novel dissipative term in the
cosmological Klein-Gordon equation.

\paragraph{Scalar field dependent dissipation function.}

Let's assume now that $Q=Q\left( \phi (t)\right) $. %Then $\frac{d}{dt}\int
%H(t)Q(\phi (t))dt=H(t)Q(\phi (t))$, and $\frac{d}{d\phi }\int H(t)Q(\phi
%(t))dt=\int H(t)Q^{\prime }(\phi (t))dt$,
In the following by a prime we denote the
derivative with respect to $\phi $, or, more generally, with the argument of
the function. In this, case the Euler-Lagrange equation
%becomes,
%\begin{eqnarray}
%&&e^{3\int H(t)Q(\phi (t))dt}\Bigg[3a^{2}\dot{a}\frac{d\phi }{dt}%
%+3a^{3}H(t)Q(\phi (t))\frac{d\phi }{dt}+a^{3}\frac{d^{2}\phi }{dt^{2}}
%\notag \\
%&&+\frac{dV\left( \phi \right) }{d\phi }a^{3}+3a^{3}\int H(t)Q^{\prime
%}(\phi (t))dt\left( \frac{1}{2}\dot{\varphi}^{2}-V\left( \phi \right)
%\right) \Bigg]=0,\nonumber\\
%\end{eqnarray}%
leads to the evolution equation of the dissipative scalar field as given
by,
\begin{eqnarray}
\hspace{-0.5cm} \ddot{\phi}&+&3H(1+Q(\phi (t)))\dot{\phi}+V^{\prime }\left(
\phi \right)  \notag \\
\hspace{-0.5cm} &-&3\int H(t)Q^{\prime }(\phi (t))dt\left( \frac{1}{2}\dot{%
\varphi}^{2}-V\left( \phi \right) \right) =0.
\end{eqnarray}

\paragraph{Hamiltonian formulation for the time dependent dissipation
function.}

We introduce now the generalized momentum, defined, in the case of the
Lagrangian (\ref{L1}) as
\begin{equation}
P_{\phi }=\frac{\partial L_{\phi }}{\partial \dot{\phi}}=a^{3}e^{3\int
H(t)Q(t)dt}\dot{\phi},
\end{equation}%
which allows to introduce the generalized, time dependent Hamiltonian
function $\mathcal{H}$ as
\begin{equation}
\mathcal{H}=P_{\phi }\dot{\phi}-L_{\phi }=a^{3}\left( \frac{1}{2}\dot{\phi}%
^{2}+V(\phi )\right) e^{3\int H(t)Q(t)dt},
\end{equation}%
as well as the generalized effective energy density of the scalar field $%
\rho _{\phi }$, defined according to

\begin{equation}
\rho _{\phi }^{(eff)}=\left( \frac{1}{2}\dot{\phi}^{2}+V(\phi )\right)
e^{3\int H(t)Q(t)dt}=\rho _{\phi }e^{3\int H(t)Q(t)dt}.  \label{rhoeff}
\end{equation}

These expressions for the Hamiltonian and energy density are also valid for
the case of the scalar field dependent dissipation function, $Q=Q(\phi (t))$.

\paragraph{Potentials depending on the time derivative of the scalar field
only.}

We consider now scalar field models in which the potential depends on the
time derivatives of the scalar field only, $V=V\left( \dot{\phi}\right) $.
Then the Euler-Lagrange equation takes the form,
\begin{equation}
\frac{d}{dt}\left[ a^{3}e^{3\int H(t)Q(t)dt}\left( \dot{\phi}-\frac{dV\left(
\dot{\phi}\right) }{d\dot{\phi}}\right) \right] =0,
\end{equation}%
and it admits the first integral,
\begin{equation}
\dot{\phi}-\frac{dV\left( \dot{\phi}\right) }{d\dot{\phi}}=\frac{C}{a^{3}}%
e^{-3\int H(t)Q(t)dt},
\end{equation}%
where $C$ is an arbitrary constant of integration. In the particular case
\begin{equation}
V\left( \dot{\phi}\right) =\left( 1-\alpha \right) \frac{\dot{\phi}^{2}}{2},
\end{equation}%
the dissipative scalar field satisfies the first order differential
equation, given by,
\begin{equation}
\dot{\phi}=\frac{C}{\alpha }\frac{e^{-3\int H(t)Q(t)dt}}{a^{3}}.
\end{equation}

\subsection{Covariant formulation of the dissipative Klein-Gordon equation}

We introduce now the Lagrangian of the dissipative scalar field in the
general covariant form as,
\begin{eqnarray}
\hspace{-0.5cm}S_{\phi } &=&\int L_{\phi }\sqrt{-g}d^{4}x  \notag  \label{L2}
\\
\hspace{-0.5cm} &=&\int e^{\Gamma \left( g_{\alpha \beta },\phi ,x^{\alpha
}\right) }\left[ \frac{1}{2}g^{\mu \nu }\nabla _{\mu }\phi \nabla _{\nu
}\phi -V\left( \phi \right) \right] \sqrt{-g}d^{4}x,  \notag \\
&&
\end{eqnarray}%
where the dissipation exponent $\Gamma \left( g_{\alpha \beta },\phi
,x^{\alpha }\right) $ is an arbitrary scalar function of the metric tensor,
of the scalar field, and of the coordinates. A particular, and useful,
representation of the dissipation function is given by the expression,
\begin{equation}
\Gamma \left( g_{\alpha \beta },\phi ,x^{\alpha }\right) =\int \nabla
_{\lambda }u^{\lambda }Q\left( \phi ,x^{\mu }\right) \sqrt{-g}d^{4}x,
\label{G1}
\end{equation}%
with $u^{\lambda }$ denoting the velocity four-vector of the cosmological
fluid. In the case of the FLRW geometry, in the comoving frame $u^{\lambda
}=\left( 1,0,0,0\right) $, and $\nabla _{\lambda }u^{\lambda }=\left( 1/%
\sqrt{-g}\right) \partial _{\alpha }\left( \sqrt{-g}u^{\alpha }\right)
=\left( 1/a^{3}\right) \frac{d}{dt}a^{3}=3H$. In an arbitrary coordinate
system \cite{LaLi},

\begin{equation}
\nabla _{\lambda }u^{\lambda }=\frac{1}{\sqrt{-g}}\frac{\partial }{\partial
x^{\lambda }}\left( \sqrt{-g}u^{\lambda }\right) =\frac{\partial u^{\lambda }%
}{\partial x^{\lambda }}+u^{\lambda }\frac{1}{\sqrt{-g}}\frac{\partial }{%
\partial x^{\lambda }}\sqrt{-g}.  \label{grad}
\end{equation}

The Euler-Lagrange equations corresponding to the action (\ref{L2}) are
given by
\begin{eqnarray}
\hspace{-0.5cm} &&\nabla _{\alpha }\left( \sqrt{-g}g^{\alpha \beta
}e^{\Gamma \left( g_{\alpha \beta },\phi ,x^{\alpha }\right) }\nabla _{\beta
}\phi \right)  \notag \\
\hspace{-0.5cm} &&+\left[ \frac{dV\left( \phi \right) }{d\phi }-\frac{%
\partial \Gamma \left( g_{\alpha \beta },\phi ,x^{\alpha }\right) }{\partial
\phi }\right] e^{\Gamma \left( g_{\alpha \beta },\phi ,x^{\mu }\right) }%
\sqrt{-g}=0,
\end{eqnarray}%
or, equivalently,
\begin{eqnarray}
&&\frac{\partial }{\partial x^{\alpha }}\left( \sqrt{-g}g^{\alpha \beta }%
\frac{\partial \phi }{\partial x^{\beta }}\right) +\sqrt{-g}g^{\alpha \beta }%
\frac{\partial \phi }{\partial x^{\beta }}\frac{\partial \Gamma \left(
g_{\alpha \beta },\phi ,x^{\alpha }\right) }{\partial x^{\alpha }}  \notag \\
&&+\left[ \frac{dV\left( \phi \right) }{d\phi }-\frac{\partial \Gamma \left(
g_{\alpha \beta },\phi ,x^{\alpha }\right) }{\partial \phi }\right] \sqrt{-g}%
=0,
\end{eqnarray}%
finally giving the dissipative covariant Klein-Gordon equation as,
\bea\label{Scalmot}
\square \phi &+&g^{\alpha \beta }\nabla _{\alpha }\phi \nabla _{\beta }\Gamma
\left( g_{\alpha \beta },\phi ,x^{\alpha }\right) +V^{\prime }(\phi )\nonumber\\
&-&\frac{%
\partial \Gamma \left( g_{\alpha \beta },\phi ,x^{\alpha }\right) }{\partial
\phi }=0.
\eea

\subsection{The energy-momentum tensor of the dissipative scalar field.}

By taking into account that $\delta \sqrt{-g}=-(1/2)\sqrt{-g}g_{\alpha \beta
}\delta g^{\alpha \beta }$, and that the Lagrangian density of the scalar
field does not depend on the derivatives of the metric, we obtain for the
energy-momentum tensor the general expression,
\begin{equation}
^{(\phi )}T_{\alpha \beta }=2\frac{\delta L_{\phi }}{\delta g^{\alpha \beta }%
}-L_{\phi }g_{\alpha \beta },
\end{equation}%
from which one obtains the energy-momentum tensor of the dissipative scalar
field as,
\begin{eqnarray}
^{(\phi )}T_{\alpha \beta } &=&e^{\Gamma \left( g_{\alpha \beta },\phi
,x^{\alpha }\right) }\Bigg[\nabla _{\alpha }\phi \nabla _{\beta }\phi
+\left( \Theta _{\alpha \beta }-g_{\alpha \beta }\right)  \notag  \label{EMT}
\\
&&\times \left( \frac{1}{2}g^{\mu \nu }\nabla _{\mu }\phi \nabla _{\nu }\phi
-V\left( \phi \right) \right) \Bigg],
\end{eqnarray}%
where we have denoted,
\begin{equation}
\Theta _{\alpha \beta }\left( g_{\mu \nu },\phi ,x^{\mu }\right) =2\frac{%
\delta \Gamma \left( g_{\alpha \beta },\phi ,x^{\mu }\right) }{\delta
g^{\alpha \beta }}.
\end{equation}

We may call the tensor $\Theta _{\alpha \beta }\left( g,\phi ,x^{\mu
}\right) $ the dissipation tensor of the scalar field. For $\Theta _{\alpha
\beta }=0$, and $\Gamma =0$, we recover the energy-momentum tensor of the ideal scalar field.

With the help of the energy density and pressure of the ideal scalar field, the energy-momentum tensor of the dissipative scalar field can be written as,
\be
^{(\phi )}T_{\alpha \beta } =e^{\Gamma \left( g_{\alpha \beta },\phi
,x^{\alpha }\right) }\left[\left(\rho_\phi+p_\phi\right)U_\alpha U_\beta+\left(\Theta _{\alpha \beta}-g_{\alpha \beta}p_\phi\right)\right].
\ee

\subsubsection{The particular case $\Gamma \left( g_{\alpha %
\beta },x^{\alpha }\right) =\int \nabla _{%
\lambda }u^{\lambda }Q\left( x^{\mu }\right) \sqrt{-g%
}d^{4}x$.}

For the particular case of the dissipation exponent given by Eq. (\ref{G1}),
for the variation with respect to the metric of $\Gamma \left( g_{\alpha
\beta },x^{\mu }\right) $ we obtain,
\begin{eqnarray}
\hspace{-0.5cm} &&\frac{\delta \Gamma \left( g_{\alpha \beta },x^{\mu
}\right) }{\delta g^{\alpha \beta }}=\frac{\delta }{\delta g^{\alpha \beta }}%
\int \nabla _{\lambda }u^{\lambda }Q\left( x^{\mu }\right) \sqrt{-g}d^{4}x
\notag \\
\hspace{-0.5cm} &=&\int \frac{\delta }{\delta g^{\alpha \beta }}\left[
\nabla _{\lambda }u^{\lambda }Q\left( x^{\mu }\right) \sqrt{-g}\right] d^{4}x
\notag \\
\hspace{-0.5cm} &=&\int \left[ \frac{\delta \left( \nabla _{\lambda
}u^{\lambda }\right) }{\delta g^{\alpha \beta }}\sqrt{-g}+\nabla _{\lambda
}u^{\lambda }\frac{\delta \sqrt{-g}}{\delta g^{\alpha \beta }}\right]
Q\left( x^{\mu }\right) d^{4}x  \notag \\
\hspace{-0.5cm} &=&\int \left[ \frac{\delta \left( \nabla _{\lambda
}u^{\lambda }\right) }{\delta g^{\alpha \beta }}+\frac{1}{\sqrt{-g}}\frac{%
\delta \sqrt{-g}}{\delta g^{\alpha \beta }}\nabla _{\lambda }u^{\lambda }%
\right] Q\left( x^{\mu }\right) \sqrt{-g}d^{4}x.  \notag \\
&&
\end{eqnarray}

With the use of Eq. (\ref{grad}), we obtain,
\begin{equation}
\frac{\delta \left( \nabla _{\lambda }u^{\lambda }\right) }{\delta g^{\alpha
\beta }}=\frac{\partial }{\partial x^{\lambda }}\frac{\delta u^{\lambda }}{%
\delta g^{\alpha \beta }}+\frac{\delta }{\delta g^{\alpha \beta }}\left(
u^{\lambda }\frac{\partial \ln \sqrt{-g}}{\partial x^{\lambda }}\right) .
\label{V1}
\end{equation}

The variation of the four-velocity with respect to the metric is given by,
\begin{equation}
\delta u^{\lambda }=\frac{1}{2}u^{\lambda }u_{\alpha }u_{\beta }\delta
g^{\alpha \beta },
\end{equation}%
which can be obtained from the relations $\delta g^{\alpha \beta }u_{\alpha
}u_{\beta }=2u_{\mu }\delta u^{\mu }$ $=2u_{\mu }\left( u^{\mu }\delta
g^{\alpha \beta }u_{\alpha }u_{\beta }/2\right) $, respectively, where we
have also used the condition of the normalization of the four-velocity, $%
g^{\alpha \beta }u_{\alpha }u_{\beta }=1$. Thus, we immediately find,
\begin{equation}
\frac{\delta u^{\lambda }}{\delta g^{\alpha \beta }}=\frac{1}{2}u^{\lambda
}u_{\alpha }u_{\beta }.
\end{equation}

Hence, Eq. (\ref{V1}) becomes,
\begin{eqnarray}
&&\frac{\delta \left( \nabla _{\lambda }u^{\lambda }\right) }{\delta
g^{\alpha \beta }}=\frac{\partial }{\partial x^{\lambda }}\frac{\delta
u^{\lambda }}{\delta g^{\alpha \beta }}+\frac{\delta u^{\lambda }}{\delta
g^{\alpha \beta }}\frac{\partial \ln \sqrt{-g}}{\partial x^{\lambda }}
\notag \\
&&+u^{\lambda }\frac{\partial }{\partial x^{\lambda }}\frac{\delta \ln \sqrt{%
-g}}{\delta g^{\alpha \beta }}=\frac{1}{2}\frac{\partial u^{\lambda }}{%
\partial x^{\lambda }}u_{\alpha }u_{\beta }+\frac{1}{2}u^{\lambda }\frac{%
\partial }{\partial x^{\lambda }}\left( u_{\alpha }u_{\beta }\right)  \notag
\\
&&+\frac{1}{2}u^{\lambda }u_{\alpha }u_{\beta }\frac{\partial \ln \sqrt{-g}}{%
\partial x^{\lambda }}-\frac{1}{2}u^{\lambda }\frac{\partial }{\partial
x^{\lambda }}g_{\alpha \beta }  \notag \\
&=&\frac{1}{2}\left( \frac{\partial u^{\lambda }}{\partial x^{\lambda }}%
+u^{\lambda }\frac{\partial \ln \sqrt{-g}}{\partial x^{\lambda }}\right)
u_{\alpha }u_{\beta }+\frac{1}{2}u^{\lambda }\frac{\partial \left( u_{\alpha
}u_{\beta }-g_{\alpha \beta }\right) }{\partial x^{\lambda }}  \notag \\
&=&\frac{1}{2}\nabla _{\lambda }u^{\lambda }u_{\alpha }u_{\beta }+\frac{1}{2}%
u^{\lambda }\frac{\partial }{\partial x^{\lambda }}\left( u_{\alpha
}u_{\beta }-g_{\alpha \beta }\right) ,
\end{eqnarray}%
\newline
where we have used the relation $\delta \ln \sqrt{-g}/\delta g^{\alpha \beta
}=\left( 1/\sqrt{-g}\right) \delta \sqrt{-g}/\delta g^{\alpha \beta }$.

Consequently,
\begin{eqnarray}
\frac{\delta \left( \nabla _{\lambda }u^{\lambda }\right) }{\delta
g^{\alpha \beta }}&+&\frac{1}{\sqrt{-g}}\frac{\delta \sqrt{-g}}{\delta
g^{\alpha \beta }}\nabla _{\lambda }u^{\lambda }  \nonumber\\
&=&\frac{1}{2}\left[ \nabla _{\lambda }u^{\lambda }h_{\alpha \beta
}+u^{\lambda }\frac{\partial }{\partial x^{\lambda }}h_{\alpha \beta }\right]
,
\end{eqnarray}
where we have introduced the projection operator $h_{\alpha \beta }$,
defined according to $h_{\alpha \beta }=u_{\alpha }u_{\beta }-g_{\alpha
\beta }$. Hence, for the variation of the dissipative exponent of this
particular case, defining the dissipation tensor of the scalar field, we
obtain,
\begin{equation}
\Theta _{\alpha \beta }=\int \left[ \nabla _{\lambda }u^{\lambda }h_{\alpha
\beta }+u^{\lambda }\frac{\partial }{\partial x^{\lambda }}h_{\alpha \beta }%
\right] Q\left( x^{\mu }\right) \sqrt{-g}d^{4}x.
\end{equation}

Therefore, for a dissipation exponent $\Gamma =\int \nabla _{\lambda
}u^{\lambda }Q\left( x^{\mu }\right) \sqrt{-g}d^{4}x$, depending on the
metric, a vector field $u^{\lambda }$, and the coordinates only, the
energy-momentum tensor of the dissipative scalar field can be written as,
\begin{widetext}
\bea
^{(\phi )}T_{\alpha \beta }&=&e^{\int \nabla _{\lambda }u^{\lambda }Q\left(
x^{\mu }\right) \sqrt{-g}d^{4}x}\nonumber\\
&&\times \Bigg[\nabla _{\alpha }\phi \nabla _{\beta
}\phi +\left( \int \left[ \nabla _{\lambda }u^{\lambda }h_{\alpha \beta
}+u^{\lambda }\frac{\partial }{\partial x^{\lambda }}h_{\alpha \beta }\right]
Q\left( x^{\mu }\right) \sqrt{-g}d^{4}x-g_{\alpha \beta }\right) \left(
\frac{1}{2}g^{\mu \nu }\nabla _{\mu }\phi \nabla _{\nu }\phi -V\left( \phi
\right) \right) \Bigg].
\eea%
\end{widetext}

\paragraph{Dissipative energy-momentum tensor in FLRW geometry.}

For an FLRW Universe, the components of the energy-momentum tensor of the
dissipative scalar field become,
\begin{eqnarray}
^{(\phi )}T_{0}^{0} &=&e^{\Gamma \left( g_{\alpha \beta },t\right) }\left[
\left( 1+\Theta _{0}^{0}\right) \frac{\dot{\phi}^{2}}{2}+\left( 1-\Theta
_{0}^{0}\right) V\left( \phi \right) \right] \nonumber\\
&=&\rho _{\phi }^{(eff)},
\end{eqnarray}%
\begin{eqnarray}
-^{(\phi )}T_{i}^{i} &=&e^{\Gamma \left( g_{\alpha \beta },t\right) }\left(
1-\Theta _{i}^{i}\right) \left( \frac{1}{2}\dot{\phi}^{2}-V\left( \phi
\right) \right)  \notag \\
&=&p_{\phi }^{(eff)}\delta _{i}^{i},i=1,2,3.
\end{eqnarray}

In the particular case of the dissipation exponent given by Eq. (\ref{G1}),
by taking into account that in the FLRW geometry $h_{00}=0$, we obtain for
the $00$ component of the energy-momentum tensor the expression,
\begin{equation}
^{(\phi )}T_{0}^{0}=e^{3\int H(t)Q(t)dt}\left( \frac{1}{2}\dot{\phi}%
^{2}+V\left( \phi \right) \right) =\rho _{\phi }^{(eff)},  \label{T1}
\end{equation}

Eq. (\ref{T1}) also gives the Hamiltonian constraint for the Lagrangian of
the dissipative scalar field.

\subsection{The dissipative Klein-Gordon equation with $e^{k_{\mu %
}x^{\mu }}$ type dissipation}

We consider now an alternative Lagrangian for the description of the
dissipative scalar field, given by
\begin{eqnarray}
\hspace{-0.5cm}S_{\phi } &=&\int L_{\phi }\sqrt{-g}d^{4}x  \notag \\
\hspace{-0.5cm} &=&\int e^{k_{\mu }x^{\mu }}\left[ \frac{1}{2}g^{\alpha
\beta }\nabla _{\alpha }\phi \nabla _{\beta }\phi -V\left( \phi \right) %
\right] \sqrt{-g}d^{4}x,
\end{eqnarray}%
where $k_{\alpha }$ are the components of a constant four-vector, and $%
x^{\alpha }$, $\alpha =0,1,2,3,$ are the coordinates on the space-time
manifold. For this form of dissipation the Euler-Lagrange equations take the
form,
\begin{equation}
\frac{\partial }{\partial x^{\alpha }}\left( \sqrt{-g}g^{\alpha \beta
}e^{k_{\mu }x^{\mu }}\frac{\partial \phi }{\partial x^{\beta }}\right) +%
\frac{dV\left( \phi \right) }{d\phi }e^{k_{\mu }x^{\mu }}\sqrt{-g}=0,
\end{equation}%
giving the following dissipative Klein-Gordon equation,
\begin{equation}
\square \phi +g^{\alpha \beta }k_{\alpha }\nabla _{\beta }\phi +\frac{%
dV\left( \phi \right) }{d\phi }=0.
\end{equation}

The energy-momentum tensor of the dissipative scalar field becomes,
\begin{eqnarray}
\hspace{-0.7cm}^{(\phi )}T_{\alpha \beta } &=&e^{k_{\mu }x^{\mu }}\Bigg[%
\nabla _{\alpha }\phi \nabla _{\beta }\phi \\
\hspace{-0.7cm} &&-\left( \frac{1}{2}g^{\mu \nu }\nabla _{\mu }\phi \nabla
_{\nu }\phi -V\left( \phi \right) \right) \left( g_{\alpha \beta }-k_{\alpha
}x_{\beta }\right) \Bigg].
\end{eqnarray}

In the case of the FLRW geometry, with $k_\alpha =\left(k_0,0,0,0\right)$,
the action of the dissipative scalar field is given by,
\begin{equation}
L_{\phi }=a^{3}e^{3k_{0}t}\left( \frac{1}{2}\dot{\phi}^{2}-V(\phi )\right) ,
\end{equation}%
leading to the dissipative Klein-Gordon equation,

\begin{equation}
\ddot{\phi}+\left( 3H+k_{0}\right) \dot{\phi}+V^{\prime }\left( \phi \right)
=0.
\end{equation}

The components of the energy-momentum tensor of this type of dissipative
scalar fields are obtained as,
\begin{equation}
^{(\phi )}T_{0}^{0}=e^{k_{0}t}\left[ \frac{1}{2}\left( 1+k_{0}t\right) \dot{%
\phi}^{2}+\left( 1-k_{0}t\right) V\left( \phi \right) \right] =\rho _{\phi
}^{(eff)},
\end{equation}%
and
\begin{equation}
^{(\phi )}T_{i}^{i}=-e^{k_{0}t}\left( \frac{1}{2}\dot{\phi}^{2}-V\left( \phi
\right) \right) =-p_{\phi }^{(eff)},i=1,2,3,
\end{equation}%
respectively.

\subsection{Potentials depending on the gradient of the scalar field}

Finally, we consider the covariant formulation of the dissipative
Klein-Gordon equation in the presence of a potential that depends on the
magnitude of the gradients of the scalar field $X$, $V=V(X)$, with $X=\nabla
_{\alpha }\phi \nabla ^{\alpha }\phi $. The action of the dissipative scalar
field system is given by
\begin{equation}
\hspace{-0.5cm}S_{\phi }=\int e^{\Gamma \left( g_{\alpha \beta },x^\mu,\phi
\right) }\left[ \frac{1}{2}g^{\alpha \beta }\nabla _{\alpha }\phi
\nabla _{\beta }\phi -V\left( X\right) \right] \sqrt{-g}d^{4}x,
\end{equation}%
leading to the Euler-Lagrange equations,
%\begin{eqnarray}
%\hspace{-0.7cm}&&\frac{\partial }{\partial x^{\alpha }}\left[ \sqrt{-g}%
%e^{\Gamma \left( g_{\alpha \beta },\phi ,x^{\mu }\right) }g^{\alpha \beta }%
%\frac{\partial \phi }{\partial x^{\beta }}\left( 1-\frac{dV(X)}{dX}\right) %
%\right]  \notag \\
%\hspace{-0.7cm}&&-\left[ \frac{1}{2}g^{\alpha \beta }\nabla _{\alpha }\phi
%\nabla _{\beta }\phi -V\left( X\right) \right] \frac{\partial \Gamma \left(
%g_{\alpha \beta },\phi ,x^{\mu }\right) }{\partial \phi }\sqrt{-g}=0,
%\end{eqnarray}
%or,
\begin{eqnarray}
&&\left( 1-\frac{dV(X)}{dX}\right) \square \phi +g^{\alpha \beta }\nabla
_{\beta }\phi \nabla _{\alpha }\Gamma \left( g_{\alpha \beta },\phi ,x^{\mu
}\right)  \notag \\
&&-g^{\alpha \beta }\nabla _{\beta }\phi \nabla _{\alpha }\frac{dV(X)}{dX} -%
\left[ \frac{1}{2}g^{\alpha \beta }\nabla _{\alpha }\phi \nabla _{\beta
}\phi -V\left( X\right) \right]  \notag \\
&&\times \frac{\partial \Gamma \left( g_{\alpha \beta },\phi ,x^{\mu
}\right) }{\partial \phi }=0.
\end{eqnarray}

The energy-momentum tensor of the scalar field in the presence of scalar
field dependent potentials is obtained in the form
\begin{eqnarray}
&&^{(\phi )}T_{\alpha \beta } =e^{\Gamma \left( g_{\alpha \beta },\phi
,x^{\mu }\right) }\Bigg[\left( 1-2\frac{dV(X)}{dX}\right) \nabla _{\alpha
}\phi \nabla _{\beta }\phi  \notag \\
&&+\left( 2\frac{\delta \Gamma \left( g_{\alpha \beta },\phi ,x^{\mu
}\right) }{\delta g^{\alpha \beta }}-g_{\alpha \beta }\right) \left( \frac{1%
}{2}g^{\mu \nu }\nabla _{\mu }\phi \nabla _{\nu }\phi -V\left( X\right)
\right) \Bigg].  \notag \\
\end{eqnarray}

\section{The Einstein gravitational field equations}\label{sect2}

In the following we consider a gravitational model, containing, besides the
gravitational term, a nonminimally coupled dissipative scalar field, with
Lagrangian density $L_{\phi }$, and an ordinary matter term, described by
the Lagrangian $L_{m}$. Hence, the action of the present theory can be
generally written down as,
\begin{widetext}
\begin{eqnarray}
S &=&\int_{\Omega }\left[ -\frac{c^{4}}{16\pi G}R(g)+L_{\phi }+L_{m}\right]
\sqrt{-g}d^{4}x  \notag  \label{Act} \\
&=&\int_{\Omega }\left[ -\frac{c^{4}}{16\pi G}R(g)+e^{\Gamma \left(
g_{\alpha \beta },x^\alpha,\phi ,\partial_\alpha \phi\right) }\left( \frac{1}{2}g^{\mu \nu }%
\frac{\partial \phi \left( x^{\alpha }\right) }{\partial x^{\mu }}\frac{%
\partial \phi \left( x^{\alpha }\right) }{\partial x^{\nu }}-V\left( \phi
\left( x^{\alpha }\right) \right) \right) +L_{m}\right] \sqrt{-g}d^{4}x.
\end{eqnarray}%
\end{widetext}

The variation of the Ricci scalar is obtained in the following form,
\begin{eqnarray}
\delta \left( R\sqrt{-g}\right) R&=&\delta \left( R_{\mu \nu }g^{\mu \nu }%
\sqrt{-g}\right)  \notag \\
& =&\left( R_{\mu \nu }-\frac{1}{2}g_{\mu \nu }\right) \sqrt{-g}\delta
g^{\mu \nu } +g^{\mu \nu }\delta R_{\mu \nu }\sqrt{-g}.  \notag \\
\end{eqnarray}

The term $g^{\mu \nu }\delta R_{\mu \nu }$ can be written as $g^{\mu \nu
}\delta R_{\mu \nu }=\nabla _{\lambda }w^{\lambda }$, where $w^{\lambda
}=g^{\mu \nu }\delta \Gamma _{\mu \nu }^{\lambda }-g^{\mu \lambda }\delta
\Gamma _{\mu \nu }^{\nu }$, with $\Gamma _{\mu \nu }^{\lambda }$denoting the
Christoffel symbols associated to the Riemannian metric $g$. In the standard
approaches to general relativity, the boundary term $g^{\mu \nu }\delta
R_{\mu \nu }\sqrt{-g}$ is canceled out with the use of the Gauss' theorem,

\begin{equation}
\int_{\Omega }g^{\mu \nu }\delta R_{\mu \nu }\sqrt{-g}d^{4}x=\int_{\Omega
}\nabla _{\lambda }w^{\lambda }\sqrt{-g}d^{4}x=\int_{\partial \Omega
}w^{\lambda }dS_{\lambda },
\end{equation}%
where $dS_{\lambda }$ is the element of integration over the hypersurface
surrounding the four-volume element $d\Omega $, under the assumption that
the variations of the field cancel at the integration limits. Hence, the
gravitational field equations in the presence of a dissipative scalar field,
and a vanishing boundary term take the form,

\begin{equation}
R_{\mu \nu }-\frac{1}{2}g_{\mu \nu }=\frac{8\pi G}{c^{4}}\left( ^{(\phi
)}T_{\mu \nu }+^{(m)}T_{\mu \nu }\right) ,
\end{equation}%
where the energy-momentum tensor of the dissipative scalar field is given by
Eq. (\ref{EMT}), while $^{(m)}T_{\mu \nu }$ is the energy-momentum tensor of
ordinary matter, defined as $^{(m)}T_{\mu \nu }=\left( 2/\sqrt{-g}\right)
\delta \left( \sqrt{-g}L_{m}\right) /\delta g^{\mu \nu }$. The variation of
the action (\ref{Act}) with respect to the scalar field $\phi $ gives the
equation of motion of the scalar field, Eq. (\ref{Scalmot}), respectively.

\subsection{The generalized Friedmann equations}

We will consider in the following the case of a dissipative scalar field
with dissipation exponent given by $\Gamma (t)=3\int H(t)Q(t)dt$. Then the
effective density $\rho _{\phi }^{(eff)}$ of the dissipative scalar field
(the Hamiltonian constraint) is given by Eq.~(\ref{rhoeff}).

We also assume that in the comoving frame the energy-momentum tensor of the
scalar field is given by $^{(\phi )}T_{0}^{0}=\rho _{\phi }^{(eff)}$, and $%
^{(\phi )}T_{i}^{i}=-p_{\phi }^{(eff)}\delta _{i}^{i}$, $i=1,2,3$. For the
adopted form of the dissipation exponent the effective energy of the scalar
field is given by Eq. (\ref{T1}).

To determine the form of the effective pressure $p_{\phi }^{(eff)}$ of the
dissipative scalar field, we impose the condition of the conservation of the
effective quantities in the cosmological background, which can be formulated
as
\begin{equation}
\dot{\rho}_{\phi }^{(eff)}+3H\left( \rho _{\phi }^{(eff)}+p_{\phi
}^{(eff)}\right) =0.  \label{Co1}
\end{equation}

Equivalently, Eq. (\ref{Co1}) can be written as,
\begin{equation}
\dot{\rho}_{\phi }+3H\left( 1+Q\right) \rho _{\phi }+3Hp_{\phi
}^{(eff)}e^{-\Gamma (t)}=0,
\end{equation}%
or,
\begin{eqnarray}
&&\dot{\phi}\ddot{\phi}+\dot{\phi}V^{\prime }(\phi )+3H(1+Q)\frac{\dot{\phi}%
^{2}}{2}+3H(1+Q)V(\phi )  \notag \\
&&+3Hp_{\phi }^{(eff)}e^{-\Gamma (t)}=0.
\end{eqnarray}

With the use of the dissipative Klein-Gordon equation (\ref{DisKG}) we can
fix now the effective form of the pressure of the dissipative scalar field
as,
\begin{equation}
p_{\phi }^{(eff)}=\left( 1+Q\right) p_{\phi }e^{\Gamma (t)}.
\end{equation}

It is easy to check that with this form of $p_{\phi }^{(eff)}$, Eq.~(\ref%
{Co1}) is equivalent to the dissipative Klein-Gordon equation (\ref{DisKG}).

Hence, for the flat FLRW metric (\ref{FLRW}), the Friedmann equations in the
presence of a dissipative scalar field, with the dissipative exponent $%
\Gamma (t)=3\int H(t)Q(t)dt$, take the form,
\begin{eqnarray}\label{Fr1}
\hspace{-0.5cm}3H^{2} &=&\frac{8\pi G}{c^{2}}\left( \rho _{\phi
}^{(eff)}+\rho _{m}c^{2}\right)  \notag\\
\hspace{-0.5cm} &=&\frac{8\pi G}{c^{2}}\left[ \left( \frac{1}{2}\dot{\phi}%
^{2}+V(\phi )\right) e^{3\int H(t)Q(t)dt}+\rho _{m}c^{2}\right] ,  \notag \\
&&
\end{eqnarray}%
\begin{eqnarray}\label{Fr2}
2\dot{H}+3H^{2} &=&-\frac{8\pi G}{c^{2}}\left( p_{\phi }^{(eff)}+p_{m}\right)
\notag\\
&=&-\frac{8\pi G}{c^{2}}\Bigg[ \left( 1+Q\right) \left( \frac{1}{2}\dot{\phi}%
^{2}-V(\phi )\right)  \notag \\
&&\times e^{3\int H(t)Q(t)dt} +p_{m}\Bigg] ,
\end{eqnarray}%
which must be considered together with the dissipative Klein-Gordon
equation, Eq. (\ref{DisKG}). By eliminating the term $3H^2$ between Eqs.~(\ref{Fr1}) and (\ref{Fr2}) we obtain the time evolution of the Hubble function as,
\bea\label{Fr3}
2\dot{H}&=&-\frac{8\pi G}{c^2}\left[\left(1+\frac{Q}{2}\right)-QV(\phi)\right]e^{3\int{H(t)Q(t)dt}}\nonumber\\
&&-\frac{8\pi G}{c^2}\left(\rho_mc^2+p_m\right).
\eea

For $Q=0$ we recover the basic equations
describing the standard quintessence cosmological models.

Once $V\left( \phi \right) $, $Q\left(t \right) $, and the equation of
state of the cosmological matter $p_{m}=p_{m}\left( \rho_{m}\right) $ are
known, the system of equations (\ref{Fr1})-(\ref{Fr3}) and (\ref{DisKG}) represents a system of
differential - integral equations for the unknowns $\left(H,\phi
,\rho_{m}\right) $.

From the Friedmann equations (\ref{Fr1}) and (\ref{Fr2}) we can obtain the
generalized conservation equation,

\begin{equation}
\frac{d}{dt}\left( a^{3}\rho _{\phi }^{(eff)}\right) +\frac{da^{3}}{dt}%
p_{\phi }^{(eff)}+\frac{d}{dt}\left( a^{3}\rho _{m}c^{2}\right) +\frac{da^{3}%
}{dt}p_{m}=0.
\end{equation}

Since we have already assumed that the effective dissipative scalar field is
conserved, it follows that the matter energy-density is also conserved, and
hence no energy-matter transfer can take place between the dissipative
scalar field, and the normal baryonic matter. Hence, the baryonic matter
content of the Universe satisfies the conservation equation,

\begin{equation}
\dot{\rho}_{m}+3H\left( \rho _{m}+\frac{p_{m}}{c^{2}}\right) =0.
\end{equation}

A useful cosmological observational quantity, the deceleration parameter $q $%
, having the definition,
\begin{equation}
q=\frac{d}{dt}\frac{1}{H}-1=-\frac{\dot{H}}{H^{2}}-1,
\end{equation}%
is obtained as,
\begin{equation}
q=\frac{1}{2}\left[ 1+3\frac{(1+Q(t))\left( \frac{1}{2}\dot{\phi}^{2}-V(\phi
)\right) e^{3\int H(t)Q(t)dt}+p_{m}}{\left( \frac{1}{2}\dot{\phi}^{2}+V(\phi
)\right) e^{3\int H(t)Q(t)dt}+\rho _{m}c^{2}}\right] .
\end{equation}

We can also introduce the parameter $w$ of the equation of state of the dark
energy, which is given by

\begin{equation}
w=\frac{p_{\phi }^{(eff)}}{\rho _{\phi }^{(eff)}}=\left( 1+Q\right) \frac{%
p_{\phi }}{\rho _{\phi }}.
\end{equation}

\subsubsection{Dimensionless form of the generalized Friedmann equations}

In order to simplify the mathematical expressions of the Friedmann
equations, we define a set of dimensionless variables $\left( \tau ,h,\Phi
,U,r_{m},P_{m}\right) $, defined according to,
\begin{eqnarray}
t &=&\frac{1}{H_{0}}\tau ,H=H_{0}h,\phi =\sqrt{\frac{3c^{2}}{8\pi G}}\Phi ,
\notag \\
V &=&\frac{3H_{0}^{2}c^{2}}{8\pi G}U,\rho _{m}=\frac{3H_{0}^{2}}{8\pi G}%
r_{m},p_{m}=\frac{3H_{0}^{2}c^{2}}{8\pi G}P_{m},
\end{eqnarray}
where $H_0$ is the present day value of the Hubble function.

The dimensionless matter density can also be written as $r_m=\rho_m/\rho_c=%
\Omega_m$, where $\rho_c=3H_0^2/8\pi G$ is the critical density, while $%
\Omega _m$ denotes the density parameter of the baryonic matter.

Then the Friedman, the Klein-Gordon, and the energy balance equations take
the form,
\begin{equation}\label{F1}
h^{2}=\left[ \frac{1}{2}\left( \frac{d\Phi }{d\tau }\right) ^{2}+U\left(
\Phi \right) \right] e^{3\int h(\tau )Q(\tau )d\tau }+r_{m},
\end{equation}%
\begin{eqnarray}  \label{F2}
2\frac{dh}{d\tau }+3h^{2} &=&-3\Bigg\{\left[ \left( 1+Q\right) \left( \frac{1%
}{2}\left( \frac{d\Phi }{d\tau }\right) ^{2}-U\left( \Phi \right) \right) %
\right]  \notag \\
&&\times e^{3\int h(\tau )Q(\tau )d\tau } +P_{m}\Bigg\},
\end{eqnarray}%
\begin{equation}
\frac{d^{2}\Phi }{d\tau ^{2}}+3h\left( 1+Q\right) \frac{d\Phi }{d\tau }%
+U^{\prime }(\Phi )=0,  \label{F3}
\end{equation}%
\begin{equation}
\frac{dr_{m}}{d\tau }+3h\left( r_{m}+P_{m}\right) =0.  \label{F4}
\end{equation}

%By eliminating the term $h^{2}$ between Eqs. (\ref{F1}) and (\ref{F2}) we
%obtain the evolution equation of $h(\tau )$ in the form,
%\begin{eqnarray}
%2\frac{dh}{d\tau }&=&-3\left[ \left( 1+\frac{Q}{2}\right) \left( \frac{d\Phi
%}{d\tau }\right) ^{2}-QU\left( \Phi \right) \right] e^{3\int h(\tau )Q(\tau
%)d\tau }  \notag \\
%&&-3\left( r_{m}+P_{m}\right) .
%\end{eqnarray}
 Moreover, we introduce the substitution,
\begin{equation}
u(\tau)=\int h(\tau )Q(\tau )d\tau ,
\end{equation}
giving $u^{\prime }(\tau)=Q(\tau)h(\tau)$. Then the system of the Friedmann-Klein-Gordon equations of the dissipative quintessence cosmology can be formulated as a second order differential system, given by,
\begin{equation}
\left( \frac{du}{d\tau }\right) ^{2}=Q^{2}\left\{ \left[ \frac{1}{2}\left(
\frac{d\Phi }{d\tau }\right) ^{2}+U\left( \Phi \right) \right]
e^{3u}+r_{m}\right\} ,
\end{equation}%
\begin{eqnarray}
&&\frac{2}{Q}\frac{d^{2}u}{d\tau ^{2}}-\frac{2}{Q^{2}}\frac{du}{d\tau }\frac{%
dQ}{d\tau }+\frac{3}{Q^{2}}\left( \frac{du}{d\tau }\right) ^{2}  \notag \\
&=&-3\left\{ \left[ \frac{1}{2}\left( \frac{d\Phi }{d\tau }\right)
^{2}-U\left( \Phi \right) \right] e^{3u}+P_{m}\right\} ,
\end{eqnarray}%
\begin{equation}
\frac{d^{2}\Phi }{d\tau ^{2}}+3\left( 1+\frac{1}{Q}\right) \frac{du}{d\tau }%
\frac{d\Phi }{d\tau }+U^{\prime }(\Phi )=0,
\end{equation}%
\begin{equation}
\frac{dr_{m}}{d\tau }+3h\left( r_{m}+P_{m}\right) =0.
\end{equation}

\subsubsection{The generalized Friedmann equations in the redshift space}

In order to allow a straightforward comparison between the theoretical
predictions and cosmological observations we introduce, instead of the time
variable, the redshift $z$, defined as $1/a=1+z$.

%Then we obtain first,
%\begin{equation}
%\frac{d}{d\tau }=\frac{dz}{d\tau }\frac{d}{dz}=-\frac{1}{a^{2}}\frac{da}{%
%d\tau }\frac{d}{dz}=-(1+z)h(z)\frac{d}{dz},
%\end{equation}%
%and,
%\begin{eqnarray}
%\frac{d^{2}}{d\tau ^{2}} &=&-\frac{dz}{d\tau }\frac{d}{dz}\left[ (1+z)h(z)%
%\frac{d}{dz}\right]  \notag \\
%&=&(1+z)h(z)\Bigg[h(z)\frac{d}{dz}  \notag \\
%&&+(1+z)h^{\prime }(z)\frac{d}{dz}+(1+z)h(z)\frac{d^{2}}{dz^{2}}\Bigg].
%\end{eqnarray}

Then the system of equations describing the cosmological evolution in the
presence of a dissipative scalar field takes the form,
\begin{equation}
\frac{du(z)}{dz}=-\frac{Q(z)}{1+z},  \label{E1}
\end{equation}%
\begin{equation}
\frac{d\Phi (z)}{dz}=-\frac{v(z)}{(1+z)h(z)},  \label{E2}
\end{equation}%
\begin{eqnarray}
h^{2}(z) &=&\left[ \frac{1}{2}\left( 1+z\right) ^{2}h^{2}(z)\left( \frac{%
d\Phi (z)}{dz}\right) ^{2}+U\left( \Phi \right) \right] e^{3u(z)}  \notag
\label{E3} \\
&&+r_{m}(z),
\end{eqnarray}%
\begin{eqnarray}
&&-2(1+z)h(z)\frac{dh(z)}{dz}+3h^{2}(z)  \notag  \label{E4} \\
&=&-3\Bigg\{\left[ \left( 1+Q(z)\right) \left( \frac{1}{2}%
(1+z)^{2}h^{2}(z)\left( \frac{d\Phi }{dz}\right) ^{2}-U\left( \Phi \right)
\right) \right]   \notag \\
&&\times e^{3u(z)}\Bigg\},
\end{eqnarray}%
\begin{equation}
-(1+z)h(z)\frac{dv(z)}{dz}+3h(z)\left( 1+Q(z)\right) v(z)+U^{\prime }(\Phi
)=0,  \label{E5}
\end{equation}%
\begin{equation}
-(1+z)\frac{dr_{m}(z)}{dz}+3r_{m}(z)=0,  \label{E6}
\end{equation}%
where we have denoted $v=d\Phi /d\tau $, and we have assumed $P_{m}=0$.

By eliminating $h^{2}(z)$ between equations (\ref{E3}) and (\ref{E4}), we
obtain for $h(z)$ the following differential equation,
\bea\label{E7}
h(z)\frac{dh(z)}{dz}&=&\Bigg[ \frac{3}{2}(1+z)h^{2}(z)\left( 1+\frac{Q(z)}{2}%
\right) \left( \frac{d\Phi (z)}{dz}\right) ^{2}\nonumber\\
&&-\frac{3}{2}\frac{Q(z)}{1+z}U(\Phi)%
\Bigg] e^{3u(z)}+\frac{3}{2}\frac{r_{m}}{1+z}.
\eea

Eqs.~(\ref{E1}-(\ref{E6}) represent a system of first order ordinary
differential equations with the unknowns $\left( u,\Phi ,v,h,r_{m}\right) $,
with the solution satisfying the constraint (\ref{E3}). In order to solve
the system, the functional form of the functions $Q(z)$ and $U(\phi )$ must
be provided. The system must be integrated with the initial conditions $%
u(0)=u_{0},\Phi (0)=\Phi _{0},v(0)=v_{0},h(0)=1$, and $r_{m}(0)=r_{m0}$,
respectively. Eq.~(\ref{E6}) can be immediately integrated to give for the
matter density parameter the expression
\begin{equation}
r_{m}(z)=\Omega _{m}(z)=\Omega _{m0}(1+z)^{3},  \label{Om}
\end{equation}%
where $\Omega _{m0}$ is the present day matter density parameter.

\section{Simple cosmological models with dissipative scalar field}\label{sect3}

In the present Section we will investigate the cosmological implications of
the dissipative scalar field models, by considering some simple analytical
form forms of the dissipation function $Q(\tau )$. We will consider the
effects of dissipation only on the late cosmological evolution, and hence we
will neglect the effects of the matter pressure in the field equations (\ref%
{E1})-(\ref{E6}), which are the basic equations describing the expansionary
dynamics of the Universe for the dissipation exponent given by $\Gamma (\tau
)=3\int {H(\tau )Q(\tau )d\tau }$.

In order to test the cosmological viability of the dissipative scalar field
model we will compare its theoretical predictions with the standard $\Lambda
$CDM model, and with a set of observational data for the Hubble function.

In a three component Universe, consisting of baryonic matter, dark matter,
and dark energy, respectively, the Hubble function of the $\Lambda $CDM
model is given by,
\begin{equation}
H=H_{0}\sqrt{\frac{\Omega _{m}^{(cr)}}{a^{3}}+\Omega _{\Lambda }}=H_{0}\sqrt{%
\Omega _{m}^{(cr)}(1+z)^{3}+\Omega _{\Lambda }},
\end{equation}%
where $\Omega _{m}^{(cr)}=\Omega _{b}^{(cr)}+\Omega _{DM}^{(cr}$, and $%
\Omega _{b}^{(cr)}=\rho _{b}/\rho _{cr}$, $\Omega _{DM}=\rho _{DM}/\rho
_{cr} $ and $\Omega _{\Lambda }=\Lambda /\rho _{cr}$ denote the density
parameters of the baryonic matter, dark matter, and dark energy,
respectively. In the $\Lambda $CDM model the deceleration parameter is given
by the relation,
\begin{equation}
q(z)=\frac{3(1+z)^{3}\Omega _{m}}{2\left[ \Omega _{\Lambda }+(1+z)^{3}\Omega
_{m}\right] }-1.
\end{equation}

In the following we adopt for the matter and dark energy density parameters
of the $\Lambda $CDM model the values $\Omega _{DM}=0.2589$, $\Omega
_{b}=0.0486$, and $\Omega _{\Lambda }=0.6911$, respectively \cite{1h}. Then
total matter density parameter $\Omega _{m}=\Omega _{DM}+\Omega _{b}$ has
the value $\Omega _{m}=0.3089$. For the present day value of the
deceleration parameter is given by $q(0)=-0.5381$, indicating that presently
the Universe is in an accelerating phase.

\subsection{The de Sitter solution}

As a first example of a cosmological model with dissipative scalar field we
will consider the case for which the Hubble function is a constant, $h=h_{0}=%
\mathrm{constant}$, corresponding to an exponential expansion of the
Universe, and with a deceleration parameter $q=-1$. Moreover, we assume a
vacuum Universe, with $r_{m}=0$. Then, by adding Eqs. (\ref{F1}) and (\ref%
{F2}) we obtain the relation,
\begin{equation}
h_{0}^{2}\frac{2+Q}{1+Q}e^{-\Gamma (\tau )}=2U\left( \Phi \right) .
\label{S1}
\end{equation}

This equation is identically satisfied for $Q=-2$, and $U\left( \Phi \right)
=0$. The Klein-Gordon equation becomes,
\begin{equation}
\frac{d^{2}\Phi }{d\tau ^{2}}-3h_{0}\frac{d\Phi }{d\tau }=0,
\end{equation}%
with the general solution given by,
\begin{equation}\label{PhiS}
\Phi (\tau )=\frac{C_{1}}{3h_{0}}e^{3h_{0}\tau }+C_{2},
\end{equation}%
where $C_{1}$ and $C_{2}$ are arbitrary constants of integration. We can
take $C_{2}=0$ without any loss of generality. For the dissipation exponent
we obtain the expression $\Gamma \left( \tau \right) =-6h_{0}\tau $. Hence,
in this simple model, the exponential expansion of the Universe is triggered
by the exponential increase of the scalar field, downsized by the decrease
of the dissipation exponent.

An alternative approach for obtaining de Sitter type solutions is based on directly solving the Friedmann constraint equation (\ref{F1}) for a constant $h$ and vanishing matter energy density. Then we obtain the differential equation,
\be
h_0^2e^{-\Gamma (\tau)}=\frac{1}{2}\left(\frac{d\Phi}{d\tau}\right)^2+U(\Phi),
\ee
which, once the field potential and the dissipation exponent are known, can be directly integrated to give the evolution of the scalar field $\Phi(\tau)$. For $U(\Phi)=0$, we obtain,
\be
\Phi(\tau)=\sqrt{2}h_0\int{e^{-\Gamma (\tau)/2}d\tau}+{\rm constant}.
\ee
For $\Gamma (\tau)=-6h_0\tau$, and taking the additive integration constant as zero, we obtain $\Phi (\tau)=\left(\sqrt{2}/3\right)e^{3h_0\tau}$, which allows to fix the integration constant $C_1$ in Eq.~{\ref{PhiS}) as $C_1=\sqrt{2}h_0$.

\subsubsection{de Sitter type expansion with constant dissipation function $%
Q_{0}\neq -2$}

Let's assume now that the dissipation function $Q$ takes constant values at
least on a finite time interval, so that $Q=Q_{0}=\mathrm{constant}\neq -2$.
Then, we obtain,
\begin{equation}
U\left( \Phi \left( \tau \right) \right) =\frac{h_{0}^{2}}{2}\frac{2+Q_{0}}{%
1+Q_{0}}e^{-3h_{0}Q_{0}\tau },Q_{0}\neq -2.
\end{equation}

In the limit of large $\tau $, the scalar field potential tends to zero, $%
\lim_{\tau \longrightarrow \infty }U\left( \Phi \left( \tau \right) \right)
=0$.

The Klein-Gordon equation takes the form,
\bea
\frac{d}{d\tau }\left( \frac{d\Phi }{d\tau }\right) ^{2}&+&6h_{0}\left(
1+Q_{0}\right) \left( \frac{d\Phi }{d\tau }\right) ^{2}\nonumber\\
&-&3h_{0}^{3}\frac{%
2+Q_{0}}{1+Q_{0}}e^{-3h_{0}Q_{0}\tau }=0.
\eea

A first integration leads to,
\begin{equation}
\frac{d\Phi }{d\tau }=\sqrt{C_{3}-\frac{\left( 2+Q_{0}\right) h_{0}^{2}}{%
\left( 1+Q_{0}\right) \left[ 1+6h_{0}\left( 1+Q_{0}\right) \right] }%
e^{-3h_{0}Q_{0}\tau }},  \label{Dphi}
\end{equation}%
where $C_{3}$ is an arbitrary integration constant, leading to the time
evolution of the scalar field as given by,
\begin{eqnarray}
&&\Phi \left( \tau \right) =\frac{2\sqrt{C_{3}}}{3h_{0}Q_{0}}\Bigg[\sqrt{1-%
\frac{h_{0}^{2}(2+Q_{0})e^{-3h_{0}Q_{0}t}}{C_{3}(1+Q_{0})\left[
6h_{0}(1+Q_{0})+1\right] }}  \notag \\
&&-\tanh ^{-1}\left( \sqrt{1-\frac{h_{0}^{2}(2+Q_{0})e^{-3h_{0}Q_{0}t}}{%
C_{3}(1+Q_{0})\left[ 6h_{0}(1+Q_{0})+1\right] }}\right) \Bigg],  \notag \\
&&
\end{eqnarray}%
where an integration constant has been set to zero. In this model the
dependence of the potential on the scalar field is given in a parametric
form, $U=U(\tau )$, $\Phi =\Phi (\tau )$. Both positive and negative values
of $Q_{0}$ are possible, and the evolution of the scalar field, and of its
potential, is basically determined by the dissipation constant. Hence,
depending on the numerical value of $Q_{0}$, a large number of exponentially
expanding vacuum cosmological models can be obtained. In the limit of large
times, as one can see from Eq. (\ref{Dphi}), $\Phi (\tau )\approx \sqrt{C_{3}%
}\tau $, and, even in the absence of the potential of the scalar field, the
de Sitter expansionary phase is triggered by the time derivative of the
scalar field.

\subsection{Models with dynamical Hubble function}

In the present Subsection we will consider two simple cosmological models in
the presence of a dissipative scalar field, and of a matter component. We
will consider two classes of models, under the assumptions that either the
scalar field potential, or its kinetic term can be neglected. A comparison
with the observational data, and with the standard $\Lambda$CDM model will
be also performed.

\subsubsection{Models with vanishing scalar field potential}

\begin{figure*}[htbp]
\centering
\includegraphics[width=8.0cm]{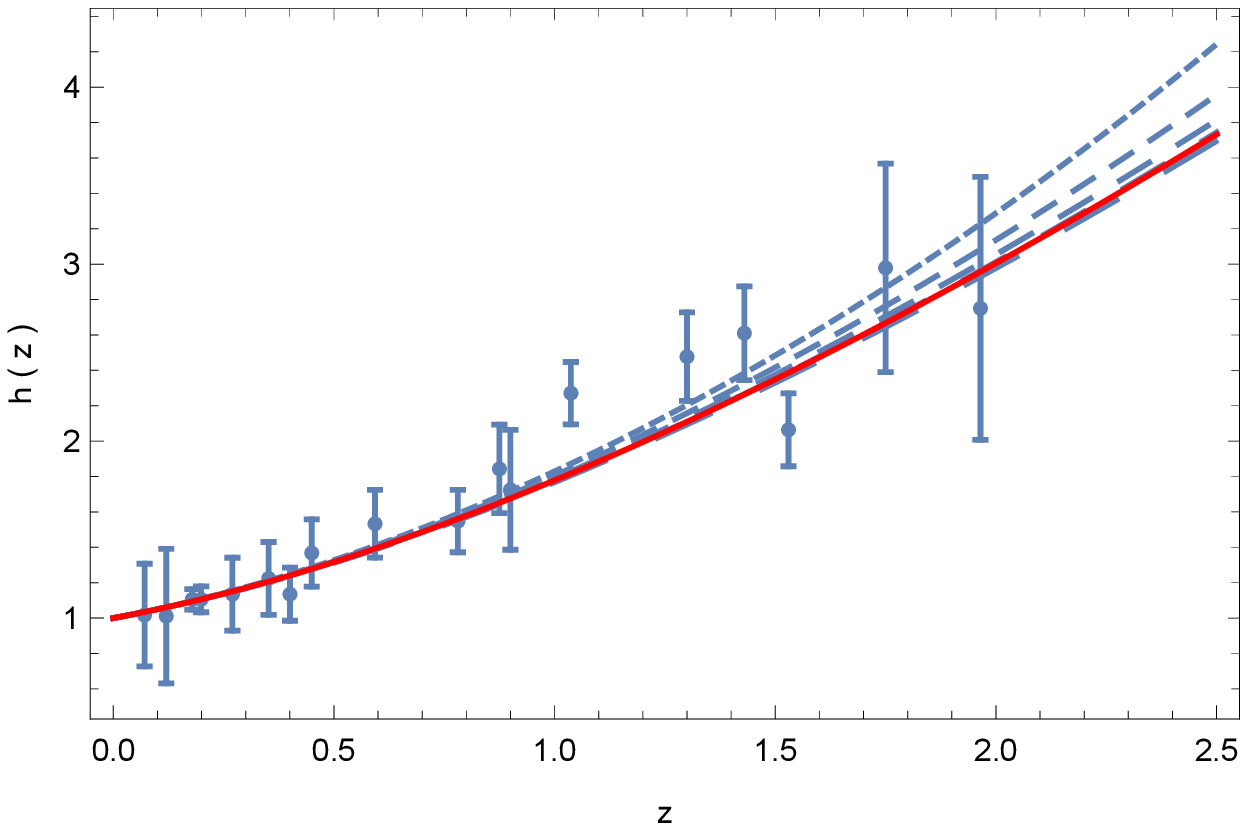} %
\includegraphics[width=8.0cm]{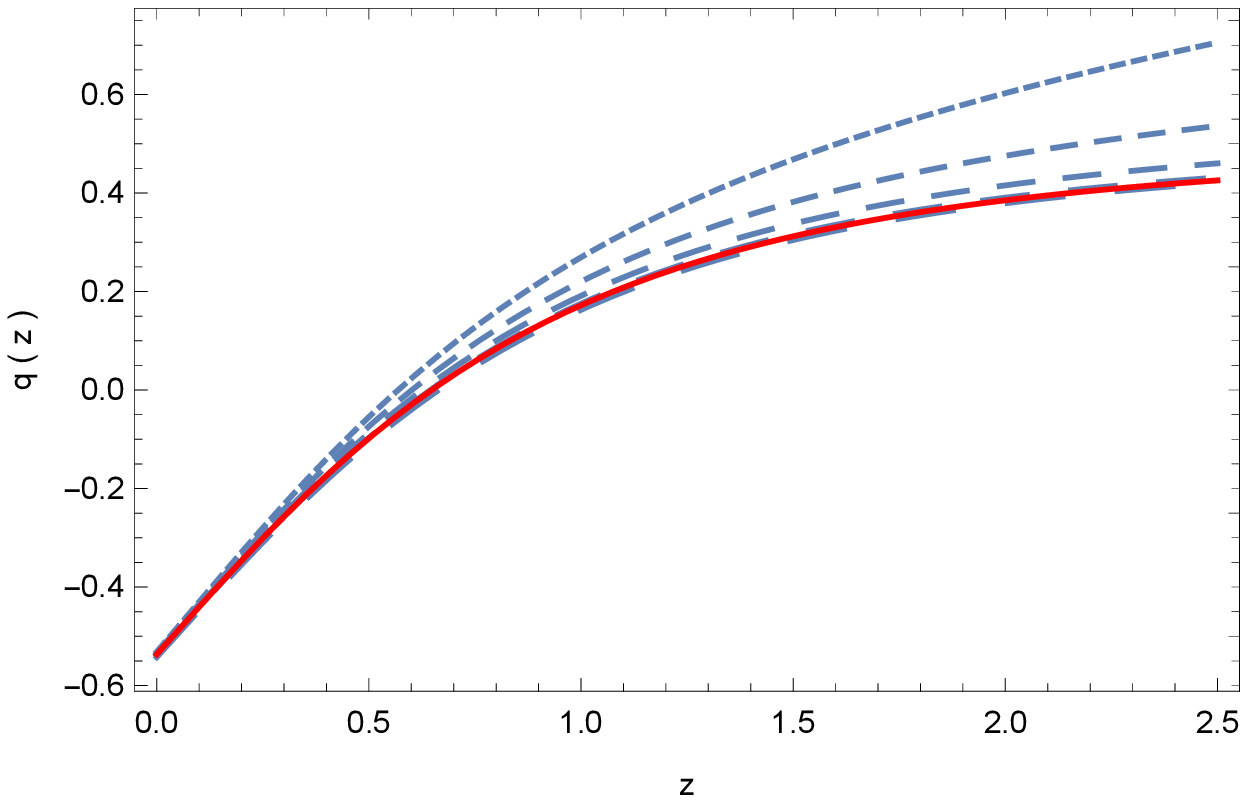}
\caption{Variation of the dimensionless Hubble function $h(z)$ (left panel),
and of the deceleration parameter $q(z)$ (right panel), in the dissipative
scalar field cosmological model with $U(\Phi)=0$, for $\Dot{\Phi}_0=0.12$, $%
\Omega_{m0}=0.30$, and for different values of $Q_0$: $Q_0=-0.29$ (dotted
curve), $Q_0=-0.49$ (short dashed curve), $Q_0=-0.69$ (dashed curve), $%
Q_0=-0.89$ (long dashed curve), and $Q_0=-1.29$ (ultra-long dashed curve),
respectively. The predictions of the $\Lambda$CDM model are represented by
the red solid curve, while the observational data are represented together
with their error bars.}
\label{fig1}
\end{figure*}

We consider now a cosmological model in the presence of a dissipative scalar
field, and of ordinary pressureless matter, in which we give up the
assumption of the global constancy of the Hubble function. For simplicity,
we assume that the dissipation function $Q$ is a constant, $Q=Q_{0}=$
constant, and that the potential of the scalar field vanishes, $U(\Phi )=0.$
Then, the evolution of the matter density is given by Eq. (\ref{Om}). From
the Klein-Gordon equation (\ref{F3}) we obtain for the time derivative of
the scalar field the expression,

\begin{equation}
\dot{\Phi}=\dot{\Phi}_{0}a^{-3\left( 1+Q_{0}\right) },
\end{equation}%
where $\dot{\Phi}_{0}$ is an arbitrary constant of integration. For the
dissipation exponent we obtain $\Gamma =3Q_{0}\int h(\tau )d\tau =3Q_{0}\ln
a $. By combining Eqs. (\ref{F1}) and (\ref{F2}), we obtain the cosmological
evolution equation of the model as,

\begin{equation}
\frac{dh}{d\tau }=-\frac{3}{2}\left( 1+\frac{Q_{0}}{2}\right) \dot{\Phi}%
^{2}e^{\Gamma }-\frac{3}{2}r_{m}.
\end{equation}

In the redshift space we obtain the following differential equation for $%
h(z) $,
\begin{eqnarray}  \label{109}
h(z)\frac{dh(z)}{dz} &=&\frac{3}{2}\left( 1+\frac{Q_{0}}{2}\right) \dot{\Phi}%
_{0}^{2}(1+z)^{5+3Q_{0}}  \notag \\
&&+\frac{3}{2}\Omega _{m0}\left( 1+z\right) ^{2}.
\end{eqnarray}

Eq.~(\ref{109}) must be integrated with the initial condition $h(0)=1$,
after fixing the numerical values of the parameters $\left( Q_{0},\dot{\Phi}%
_{0},\Omega _{m0}\right) $.

The variations of the Hubble function, and of the deceleration parameter,
are presented, as a function of the redshift $z$, in Fig.~\ref{fig1}. The
cosmological parameters corresponding to the $\Lambda$CDM model are also
shown, together with a set of observational data for the Hubble function, as
compiled in \cite{Bou}.

As one can see from Fig.~\ref{fig1}, this simple dissipative cosmological
model gives a good description of the observational data, and coincides with
the predictions of the $\Lambda $CDM model for a large range of redshifts.
The Hubble function can be expressed in an exact form as,
\begin{equation}
h(z)=\sqrt{1+\frac{\dot{\Phi}_{0}^{2}}{2}\left[ (1+z)^{3\left(
2+Q_{0}\right) }-1\right] +\Omega _{m0}\left[ (1+z)^{3}-1\right] },
\end{equation}%
while the deceleration parameter can be obtained in the form,
\begin{eqnarray}
&&q(z) =\frac{(1+z)}{h(z)}\frac{dh(z)}{dz}-1=  \notag \\
&&\frac{(z+1)\left[ 3\Omega _{m0}(1+z)^{2}+\frac{3}{2}\dot{\Phi}%
_{0}^{2}(2+Q_{0})(1+z)^{3Q_{0}+5}\right] }{2\left\{ 1+\frac{1}{2}\dot{\Phi}%
_{0}^{2}\left[ (1+z)^{3\left( 2+Q_{0}\right) }-1\right] +\Omega _{m0}\left[
(1+z)^{3}-1\right] \right\} }  \notag \\
&&-1.
\end{eqnarray}

The parameter of the equation of state of the dissipative scalar field takes
the form,
\begin{equation}
w=\left( 1+Q_{0}\right) =\mathrm{constant}.
\end{equation}

The best fit with the observational data is provided for $Q_{0}=-1.29$,
which gives $w=-0.29$. The variations of the effective density $%
\rho_{\phi}^{(eff)}$ of the scalar field, and of its effective pressure $%
p_{\rho}^{(eff)}$ are represented in Fig.~\ref{fig2}.

\begin{figure*}[htbp]
\centering
\includegraphics[width=8.0cm]{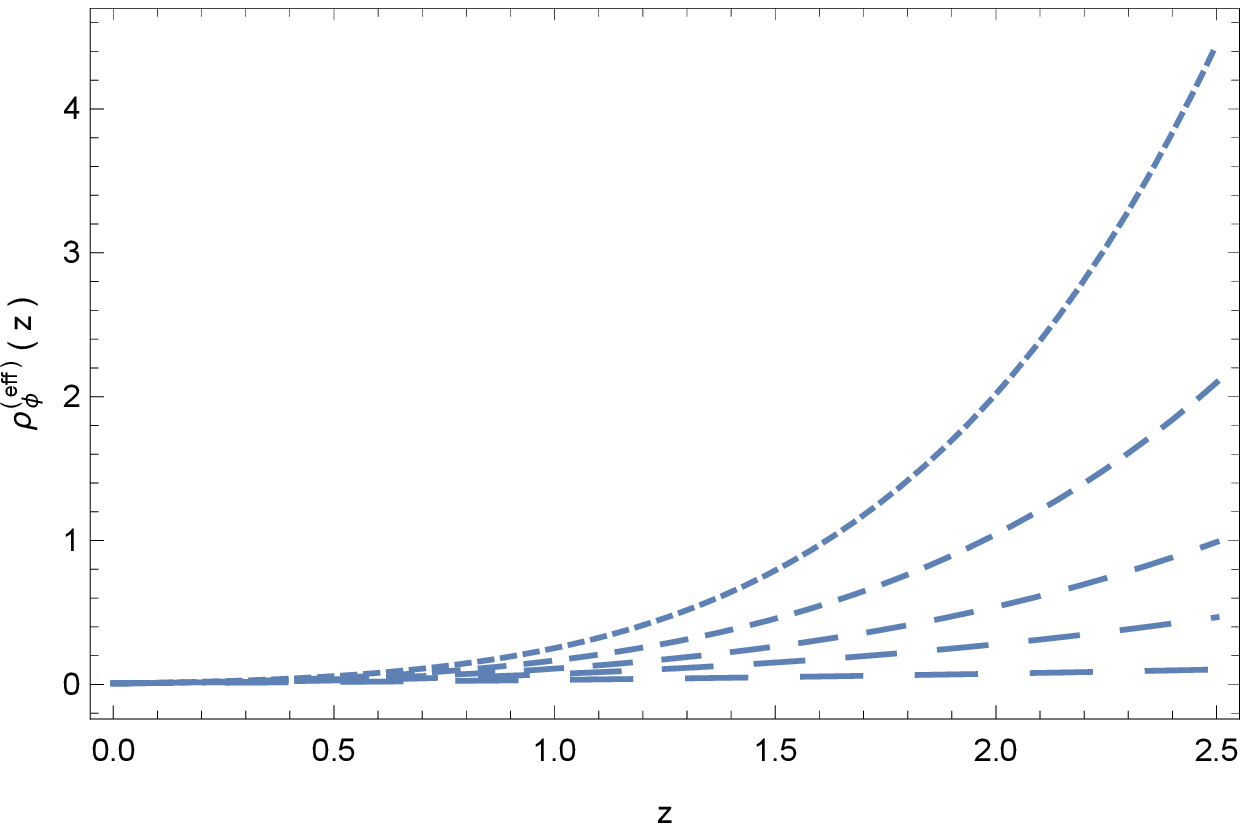} %
\includegraphics[width=8.0cm]{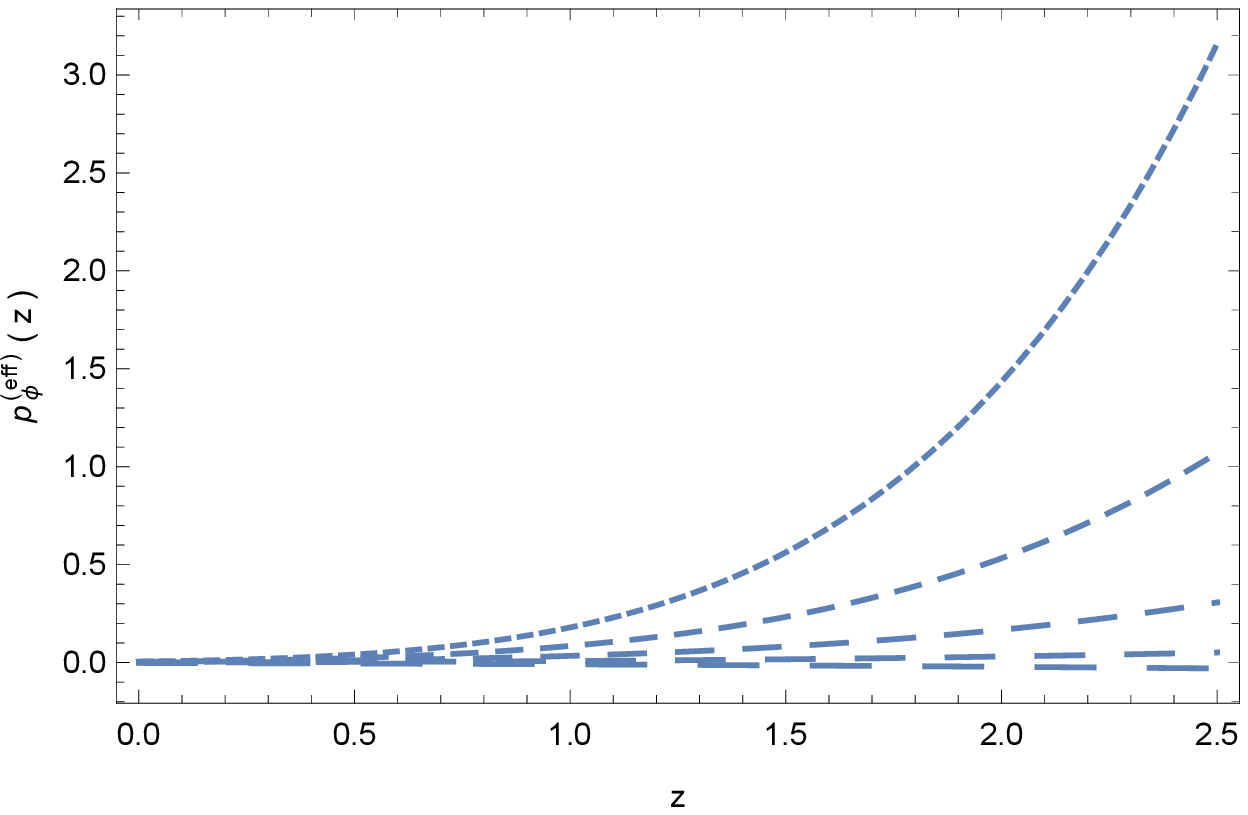}
\caption{Variation of the effective energy density $\rho_{%
\phi}^{(eff)}$ of the scalar field (left panel), and of the effective
pressure $p_{\rho}^{(eff)}$ (right panel), in the dissipative scalar
field cosmological model with $U(\Phi)=0$, for $\Dot{\Phi}_0=0.12$, $%
\Omega_{m0}=0.30$, and for different values of $Q_0$: $Q_0=-0.29$ (dotted
curve), $Q_0=-0.49$ (short dashed curve), $Q_0=-0.69$ (dashed curve), $%
Q_0=-0.89$ (long dashed curve), and $Q_0=-1.29$ (ultra-long dashed curve),
respectively.}
\label{fig2}
\end{figure*}

For the best fit values $Q_0=-1.29$, the effective density and pressure of
the scalar field are constants, with the pressure taking small negative
values. Such the dissipative scalar field behaves like a cosmological
constant even in the absence of the potential term.

For the sake of completeness, we will also consider one more parameter for
the present cosmological model, which allows testing its viability, namely,
the $Om(z)$ diagnostic, with,
\begin{equation}
Om(z)=\frac{h^{2}(z)-1}{(1+z)^{3}-1}.
\end{equation}

For the $\Lambda $CDM model, the function $Om(z)$ is a constant, equal to
the present day matter density $\Omega _{m0}$. The variation of the $Om(z)$
function is represented in Fig.~\ref{fig3}.

\begin{figure}[htbp]
\centering
\includegraphics[width=8.0cm]{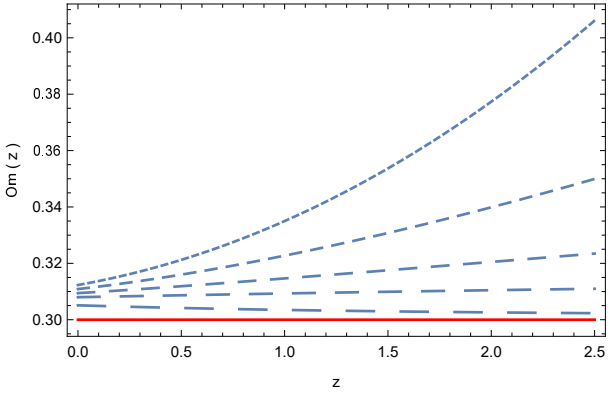}
\caption{Variation of the $Om(z)$ diagnostic function in the dissipative
scalar field cosmological model with $U(\Phi)=0$, for $\Dot{\Phi}_0=0.12$, $%
\Omega_{m0}=0.30$, and for different values of $Q_0$: $Q_0=-0.29$ (dotted
curve), $Q_0=-0.49$ (short dashed curve), $Q_0=-0.69$ (dashed curve), $%
Q_0=-0.89$ (long dashed curve), and $Q_0=-1.29$ (ultra-long dashed curve),
respectively. The solid red curve represents the $Om(z)$ function in the $%
\Lambda$CDM cosmology.}
\label{fig3}
\end{figure}

For $Q_0=-1.29$, the behavior of the $Om(z)$ function in this dissipative
scalar field model is very close to its behavior in the standard $\Lambda$%
CDM paradigm.

\subsubsection{Dissipative scalar field models with negligible kinetic term}

We consider now the case in which the potential term dominates the effective
energy density and pressure of the scalar field, that is, $U(\Phi )$
satisfies the condition $U(\Phi )>>\dot{\Phi}^{2}/2$. For simplicity, we
assume that the scalar field potential is given by the expression,
\begin{equation}
U(\Phi )=\frac{m}{2}\Phi ^{2},  \label{Potq}
\end{equation}%
where $m$ is a constant. In the following we also neglect the matter
pressure, taking $P_{m}=0$, and assume that the dissipation function is a
constant. Then, in the redshift space, the system of equations describing
the evolution of the scalar field and of the Hubble function takes the form,
\begin{equation}
\frac{d\Phi (z)}{dz}=-\frac{u(z)}{(1+z)h(z)},  \label{Ex2}
\end{equation}%
\begin{equation}
(1+z)h(z)\frac{du(z)}{dz}-3h(z)\left( 1+Q_{0}\right) u(z)-m\Phi (z)=0,
\label{Ex3}
\end{equation}%
\begin{equation}
h(z)\frac{dh(z)}{dz}=-\frac{3}{4}Q_{0}m\Phi ^{2}(z)(1+z)^{3Q_{0}-1}+\frac{3}{%
2}\Omega _{m0}\left( 1+z\right) ^{2}.  \label{Ex4}
\end{equation}

The system of equations (\ref{Ex2})-(\ref{Ex4}) must be solved with the
initial conditions $\Phi (0)=\Phi _{0}$, $u(0)=u_{0}$, and $h(0)=1$,
respectively.

The redshift evolutions of the Hubble function and of the deceleration
parameter of the dissipative scalar field model with negligible kinetic term
are represented in Fig.~\ref{fig4}. As one can see from the two panels of
Fig.~\ref{fig4}, with the values of $Q_0$ moving into the negative range,
the concordance with the cosmological data and the $\Lambda$CDM model
becomes better and better for both $h(z)$, and $q(z)$, and for $Q_0=-0.45$,
both the Hubble function and the deceleration parameter are basically
visually indistinguishable from the predictions of the standard cosmological
paradigm.

\begin{figure*}[htbp]
\centering
\includegraphics[width=8.0cm]{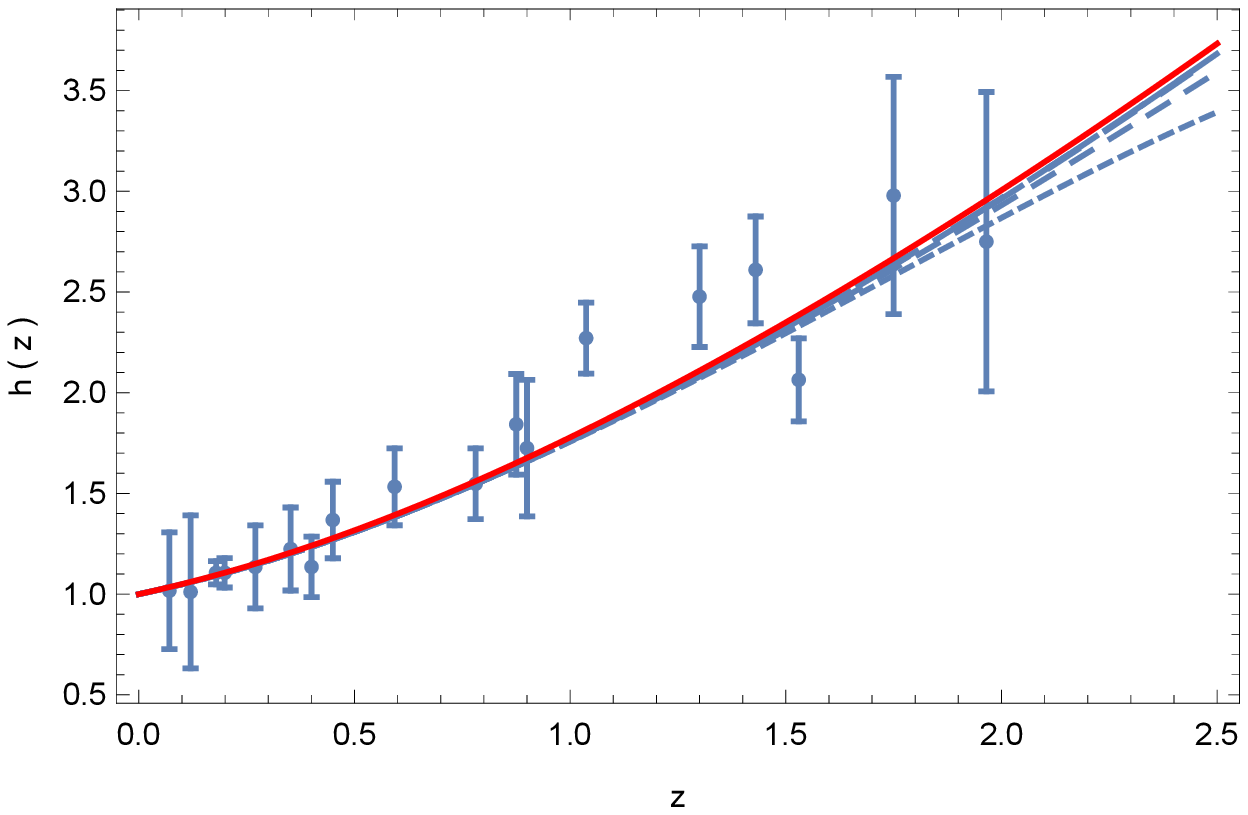} %
\includegraphics[width=8.0cm]{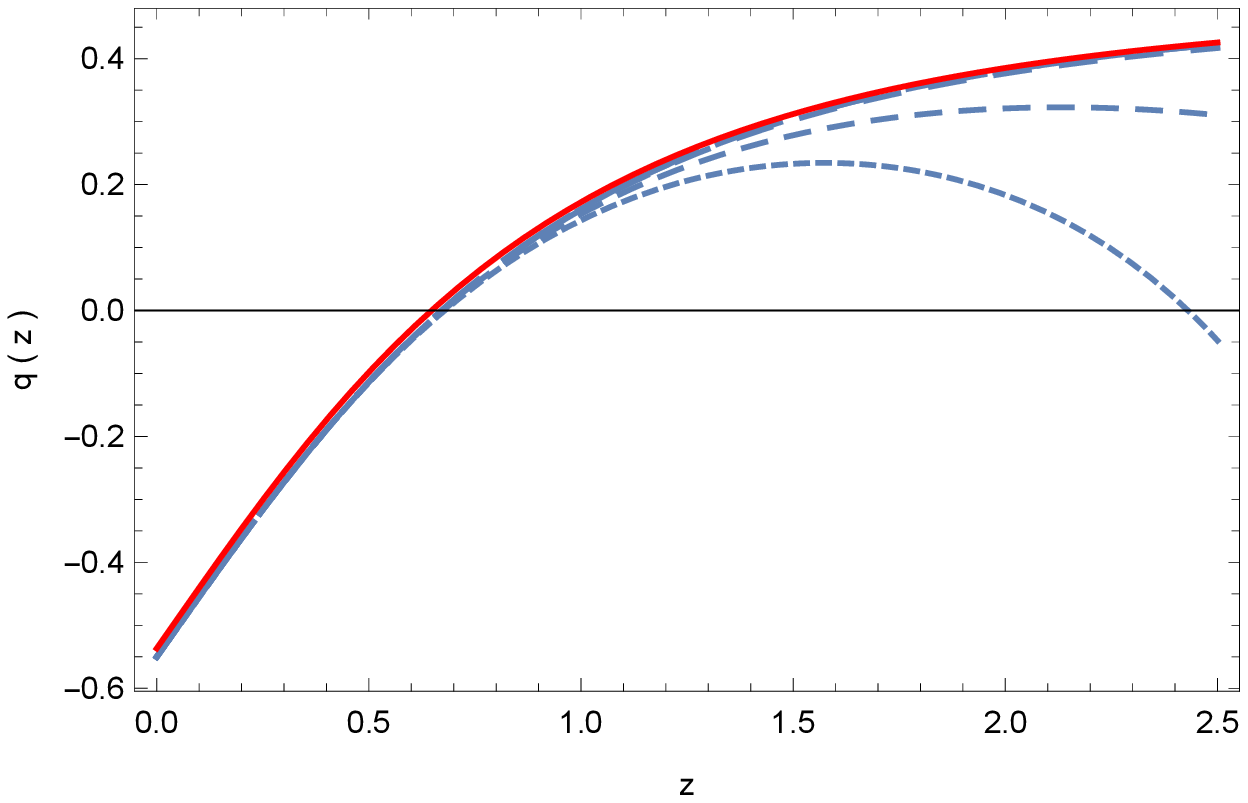}
\caption{Variation of the dimensionless Hubble function $h(z)$ (left panel),
and of the deceleration parameter $q(z)$ (right panel), in the dissipative
scalar field cosmological model with negligibly kinetic term, and $%
U(\Phi)=m\Phi^2/2$, for $\Phi (0)=0.11$, $u(0)=0.30$, $\Omega_{m0}=0.30$, $m=0.12$, and for different
values of $Q_0$: $Q_0=0.45$ (dotted curve), $Q_0=0.35$ (short dashed curve),
$Q_0=0.15$ (dashed curve), $Q_0=-0.15$ (long dashed curve), and $Q_0=-0.45$
(ultra-long dashed curve), respectively. The predictions of the $\Lambda$CDM
model are represented by the red solid curve, while the observational data
are represented together with their error bars.}
\label{fig4}
\end{figure*}

The variation of the scalar field potential, and the $Om(z)$ diagnostic
function are presented in Fig.~\ref{fig5}. The scalar field potential is
roughly a constant, almost exactly mimicking a cosmological constant. The
redshift variation of the $\Phi ^2$ type potential is (almost) exactly
compensated by the dissipation exponent, resulting in an almost constant
contribution to the Friedmann equations. However, the cosmological
evolution, even accelerated, is not exactly of the de Sitter type. The $%
Om(z) $ function also tends towards its $\Lambda$CDM value, and thus this
cosmological parameter is well recovered in the dissipative scalar field
cosmology.

The parameter of the equation of state of the dissipative quintessence type
dark energy is given by
\begin{equation}
w=-\left(1+Q_0\right)\approx -0.55,
\end{equation}
if one uses the best empirical approximation of $Q_0$. This constant
negative equation of state is different from the equation of state of the
quintessence fields with negligible kinetic terms, which is $w=-1$.

\begin{figure*}[htbp]
\centering
\includegraphics[width=8.0cm]{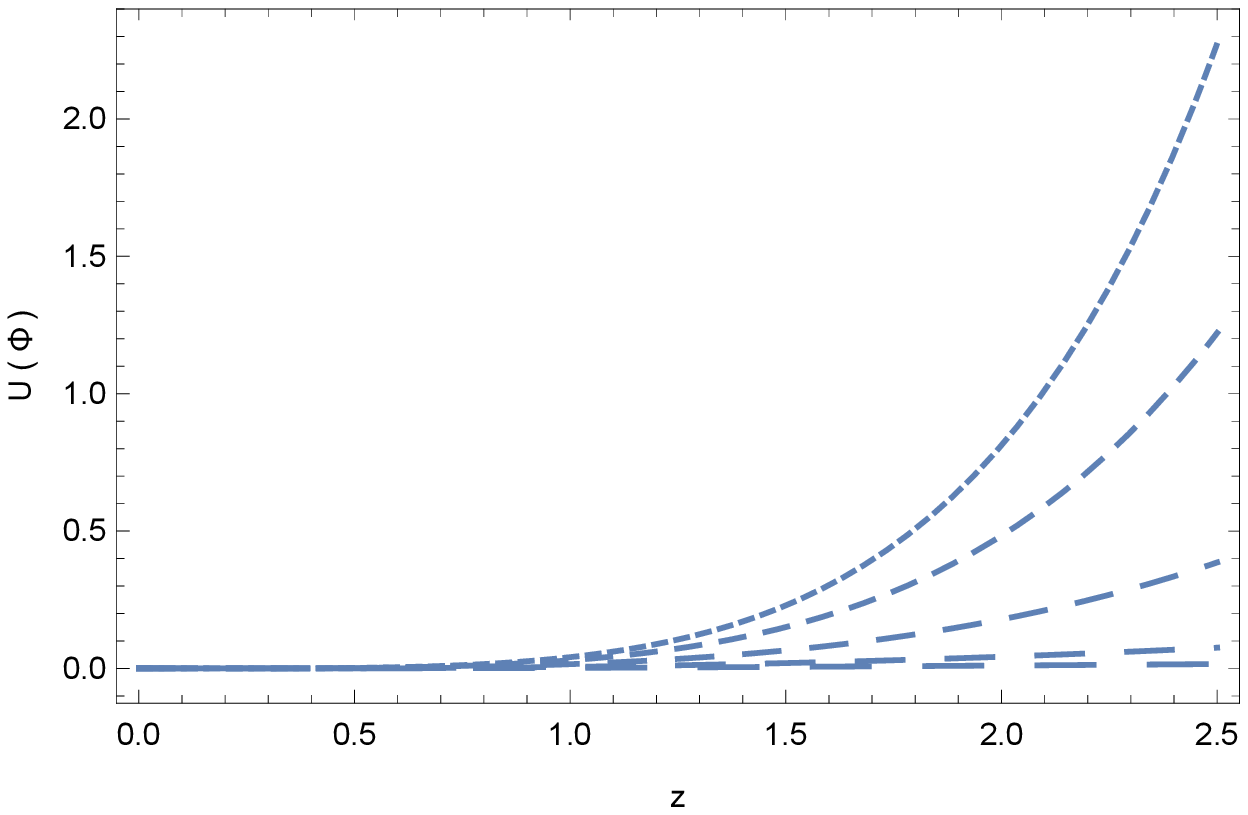} %
\includegraphics[width=8.0cm]{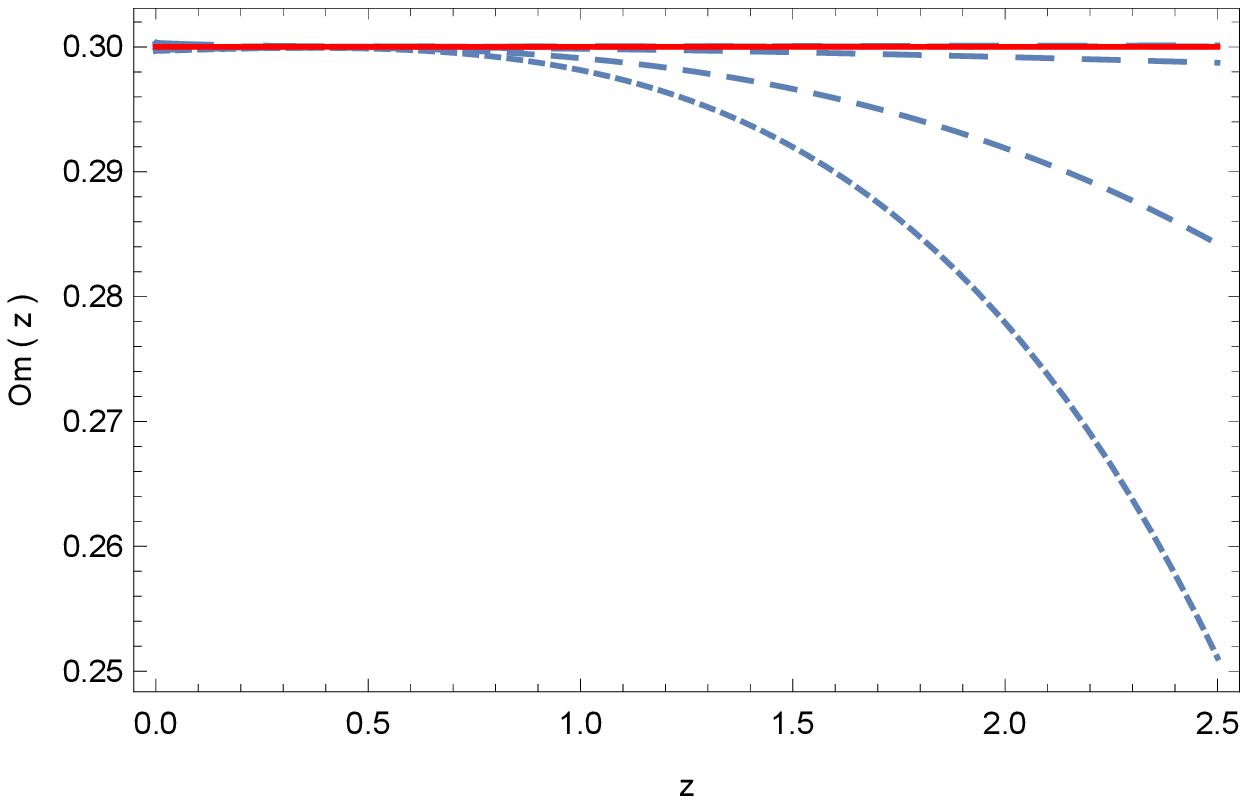}
\caption{Variation of the scalar field potential $U(\Phi)=m\Phi ^2/2$ (left
panel), and of the $Om(z)$ function (right panel), in the dissipative scalar
field cosmological model with negligibly kinetic term, for $\Phi (0)=0.11$, $%
u(0)=0.30$, $\Omega_{m0}=0.30$, $m=0.12$, and for different values of $Q_0$:
$Q_0=0.45$ (dotted curve), $Q_0=0.35$ (short dashed curve), $Q_0=0.15$
(dashed curve), $Q_0=-0.15$ (long dashed curve), and $Q_0=-0.45$ (ultra-long
dashed curve), respectively. The predictions of the $\Lambda$CDM model are
represented by the red solid curve.}
\label{fig5}
\end{figure*}

\section{Cosmological models with dynamical dissipation function}\label{sect4}

We consider now a more general class of cosmological models, in which the
dissipation function is dynamical. For simplicity, we adopt for $Q$ a simple
functional representation as,
\begin{equation}\label{134}
Q(z)=Q_{0}\left( 1+z\right) ^{\alpha },
\end{equation}%
where $Q_{0}$ and $\alpha $ are constants. For the potential of the scalar
field we still adopt the simple quadratic form (\ref{Potq}), and we also
keep the kinetic term of the field in the mathematical formalism. The system
of equations to be solved are Eqs. (\ref{E1})-(\ref{E6}), together with a
set of appropriately chosen initial condition. By taking into account the
explicit form of $Q(z)$, Eq. (\ref{E1}) can be integrated to give
\begin{equation}
u(z)=-\frac{Q_{0}}{\alpha }\left( 1+z\right) ^{\alpha }.
\end{equation}

Then, the equations describing the cosmological evolution of the Universe in
the presence of a dissipative scalar field with dynamic dissipation function
take the form,
\begin{equation}
\frac{d\Phi (z)}{dz}=-\frac{v(z)}{(1+z)h(z)},  \label{Fi1}
\end{equation}%
\begin{eqnarray}
h(z)\frac{dh(z)}{dz} &=&\frac{3}{4}\Bigg\{\left[ 2+Q_{0}\left( 1+z\right)
^{\alpha }\right] \left( 1+z\right) h^{2}[z]\left( \frac{d\Phi }{dz}\right)
^{2}  \notag  \label{Fi2} \\
&&-mQ_{0}\left( 1+z\right) ^{\alpha -1}\Phi ^{2}(z)\Bigg \}e^{-\frac{3Q_{0}}{%
\alpha }\left( 1+z\right) ^{\alpha }}  \notag \\
&&+\frac{3}{2}\Omega _{m0}\left( 1+z\right) ^{2},
\end{eqnarray}%
\bea\label{Fi3}
&&(1+z)h(z)\frac{dv(z)}{dz}-3h(z)\left[ 1+Q_{0}\left( 1+z\right) ^{\alpha }%
\right] v(z)\nonumber\\
&&\quad \quad \quad \quad \quad \quad \quad \quad-m\Phi (z)=0.
\eea

The system of equations (\ref{Fi1})-(\ref{Fi3}) must be integrated with the
initial conditions $\Phi (0)=\Phi_0$, $v(0)=v_0$, and $h(0)=1$, once the
numerical values of the parameters $\left(Q_0,\alpha, m\right)$ have been
specified.

For the sake of comparison we also present the cosmological evolution in the presence of the ideal quintessence field with quadratic potential, with $\Gamma =0$, that is, in the absence of any dissipative phenomena. The results of the numerical integration of the ideal quintessence field equations are represented by an orange curve.

The redshift variations of the Hubble function and of the deceleration
parameter are represented in Fig.~\ref{fig6}, for a constant $Q_0$ and
different values of $\alpha$. The numerical results show a relatively strong
dependence on the numerical values of the parameter $\alpha$, but for $%
\alpha =-0.60$, the predictions of the dissipative scalar field cosmological
model, and of the $\Lambda$CDM model basically coincide for both the Hubble
function and the deceleration parameter. For low redshifts, up to $z\approx
1.5$, the cosmological evolution is basically independent on the numerical
values of $\alpha$, and the concordance with the $\Lambda$CDM model is very good,
at least for the rescaled Hubble function $h(z)$. The model can also reproduce very well the predictions of the $\Lambda$CDM model for the deceleration parameter.

\begin{figure*}[htbp]
\centering
\includegraphics[width=8.0cm]{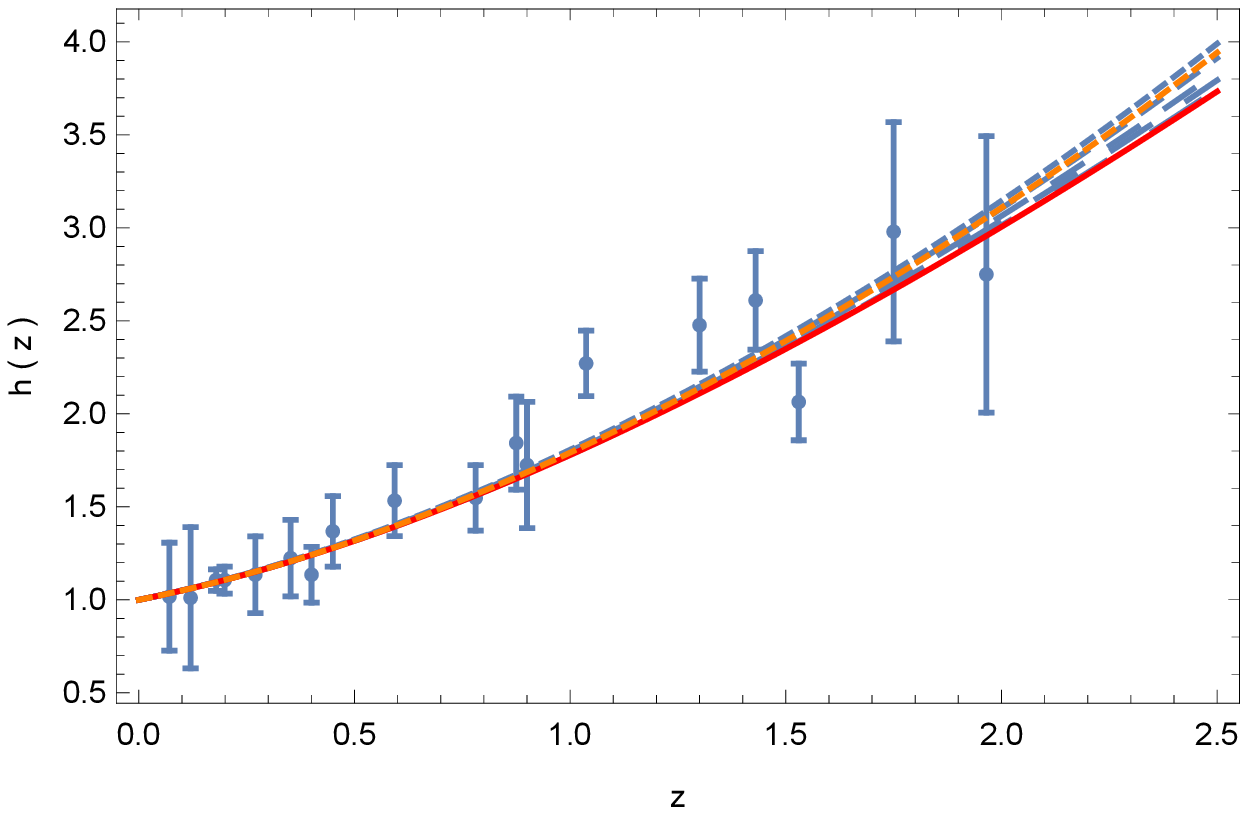} %
\includegraphics[width=8.0cm]{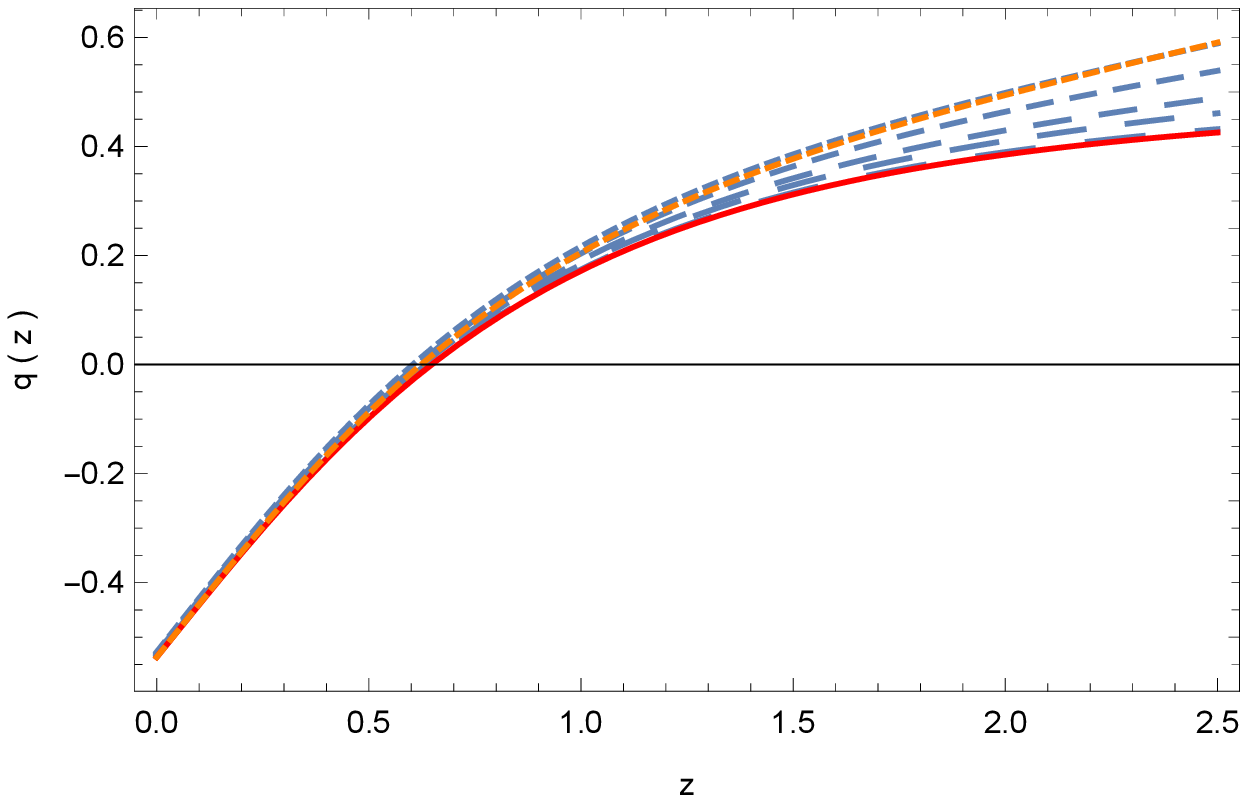}
\caption{Variation of the dimensionless Hubble function $h(z)$ (left panel),
and of the deceleration parameter $q(z)$ (right panel), in the dissipative
scalar field cosmological model with dynamical dissipation function, and
quadratic scalar field potential $U(\Phi)=m\Phi^2/2$, for $\Phi (0)=0.11$, $%
v(0)=0.30$, $\Omega_{m0}=0.3089$, $m=0.12$, $Q_0=-0.89$, and for different
values of $\alpha$: $\alpha=-1.29$ (dotted curve), $%
\alpha=-1.15$ (short dashed curve), $\alpha =-0.98$ (dashed curve), $%
\alpha=-0.85$ (long dashed curve), and $\alpha=-0.60$
(ultra-long dashed curve), respectively. The predictions of the $\Lambda$CDM
model are represented by the red solid curve, while the observational data
are given together with their error bars. The evolution of the cosmological parameters of the ideal quintessence field with quadratic potential, with $\Gamma =0$, is represented, for $m=0.682$, $\Phi (0)=0.19$, and $v(0)=0.01$ by the orange curve.  }
\label{fig6}
\end{figure*}

The variations of the effective energy density of the dissipative scalar field, as well as the behavior of the effective pressure for the quadratic  field potential, are represented in Fig.~\ref{fig7}. For the best fit values of the model with the cosmological observations both the energy density and the pressure become
approximately constant in the considered range of $z$, and hence they mimic a cosmological constant.

\begin{figure*}[tbph]
\centering
\includegraphics[width=8.0cm]{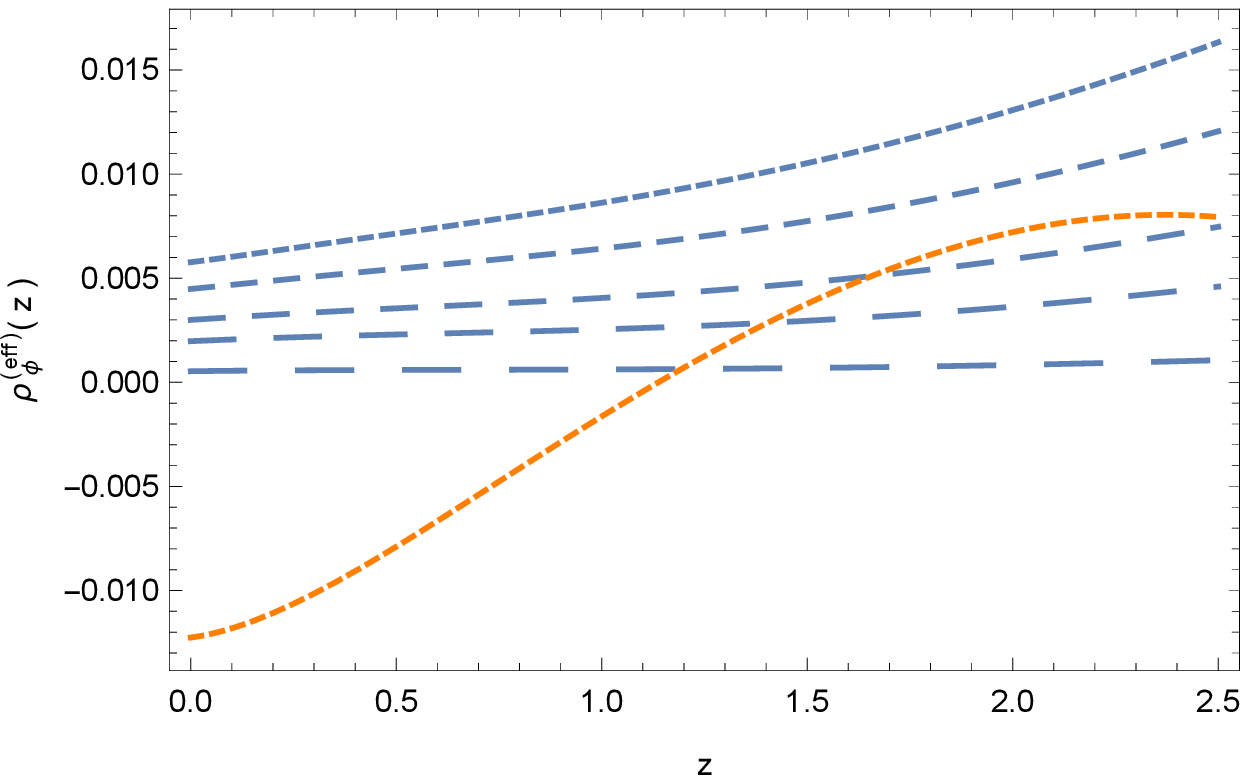} %
\includegraphics[width=8.0cm]{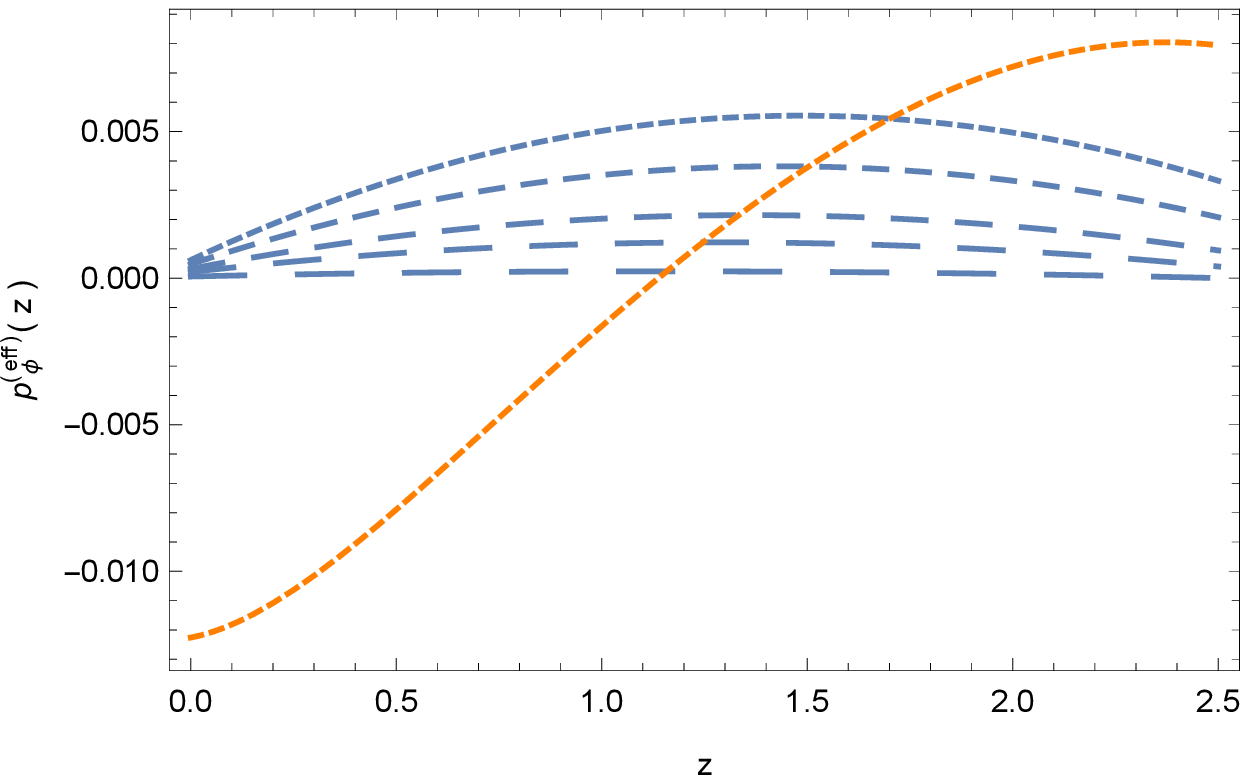}
\caption{Variation of the effective energy density of the  scalar field
(left panel), and of the effective pressure (right panel), in the dissipative
scalar field cosmological model with dynamical dissipation function, for $%
\Phi (0)=0.11$, $v(0)=0.30$, $\Omega _{m0}=0.3089$, $m=0.12$, $Q_{0}=-0.89$,
and for different values of $\alpha $: $\alpha =-1.29$
(dotted curve), $\alpha =-1.15$ (short dashed curve), $%
\alpha =-0.98$ (dashed curve), $\alpha =-0.85$ (long dashed curve),
and $\alpha =-0.60$ (ultra-long dashed curve), respectively. The evolution of the cosmological parameters of the ideal quintessence field with quadratic potential, with $\Gamma =0$, is represented, for $m=0.682$, $\Phi (0)=0.19$, and $v(0)=0.01$ by the orange curve.}
\label{fig7}
\end{figure*}

The parameter $w(z)$ of the equation of state of the scalar field is given
by,
\begin{equation}
w(z)=\frac{\left[ 1+Q_{0}(1+z)^{\alpha }\right] \left[ (1+z)^{2}h^{2}(z)%
\left( \frac{d\Phi (z)}{dz}\right) ^{2}-m\Phi ^{2}(z)\right] }{\left[
(1+z)^{2}h^{2}(z)\left( \frac{d\Phi (z)}{dz}\right) ^{2}+m\Phi ^{2}(z)\right]
}.
\end{equation}

The variation of the functions $w(z)$ and $Om(z)$ are represented in Fig.~\ref{fig8}. It is interesting to note that even the parameter of the equation of state of the scalar field is positive for all considered redshift range, the model still can explain satisfactorily the observational data, and gives almost the same predictions as the $\Lambda$CDM model.  The behavior of the $Om(z)$ function is
strongly dependent on the numerical values of $\alpha$, but for $\alpha
=-0.60$ it approaches significantly the $\Lambda$CDM value.

\begin{figure*}[tbph]
\centering
\includegraphics[width=8.0cm]{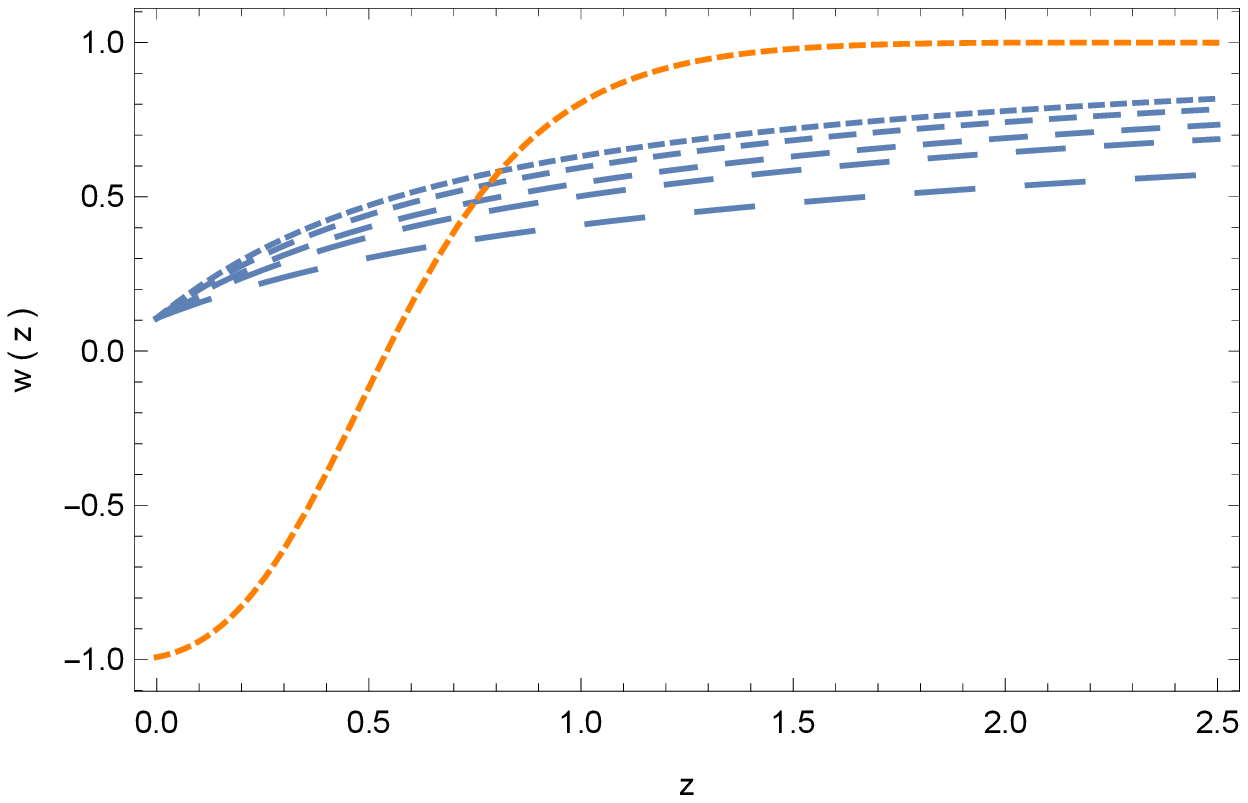}
\includegraphics[width=8.0cm]{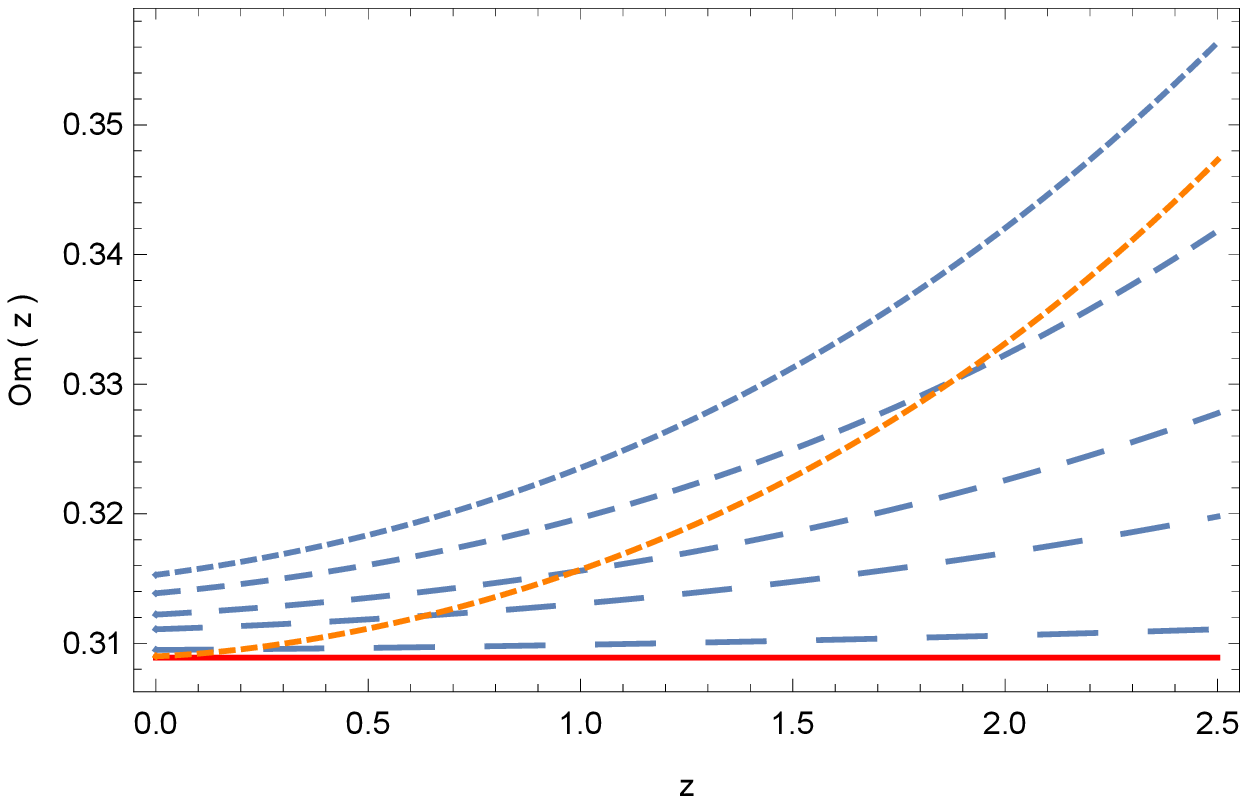}
\caption{Variation of the equation of state $w(z)$ (left panel) and of the $Om(z)$ function (right panel) in the dissipative scalar field
cosmological model with dynamical dissipation function, for $\Phi (0)=0.11$,
$v(0)=0.30$, $\Omega _{m0}=0.3089$, $m=0.12$, $Q_{0}=-0.89$, and for
different values of $\alpha $: $\alpha =-1.29$ (dotted
curve), $\alpha =-1.15$ (short dashed curve), $\alpha =-0.98$
(dashed curve), $\alpha =-0.85$ (long dashed curve), and $%
\alpha =-0.60$ (ultra-long dashed curve), respectively. The
prediction of the $\Lambda $CDM model for the $Om(z)$ function is
represented by the red solid line. The evolution of the cosmological parameters of the ideal quintessence field with quadratic potential, with $\Gamma =0$, is represented, for $m=0.682$, $\Phi (0)=0.19$, and $v(0)=0.01$ by the orange curve.}
\label{fig8}
\end{figure*}

On the other hand, as one can see from Figs.~\ref{fig6}, \ref{fig7}, and \ref{fig8}, by using a different set of values for the potential parameter $m$ and for the initial conditions $\Phi (0)$ and $v(0)$, the ideal quintessence field model with quadratic potential can also give a good description of the observational data for the Hubble function, and of the $\Lambda$CDM model. However, significant differences do appear in the behaviors of the energy density and pressure of the ideal and dissipative scalar field, as well as in the parameter of the equation of state of the dark energy.

Hence, at least in principle, it is possible to construct ideal quintessence models that mimic their dissipative counterparts at the background evolution level by adopting different values for the potential parameters, and for the initial conditions of the scalar field. The opposite situation may also be possible, with dissipative scalar field models giving an equivalent effective description of ideal quintessential field models. However, a rigorous statistical analysis of the observational datasets (Hubble, Pantheon etc.) may still allow to clearly discriminate between ideal and dissipative quintessence field models, due to their very different predictions for the parameter of the dark energy equation of state.

Nevertheless, important differences may appear at the perturbative level between ideal and dissipative quintessence models. In \cite{new2} it was shown, after performing a dynamical system analysis of the background and perturbation equations in the $\Lambda$CDM cosmology and in the quintessence models with an exponential potential, that in the case of quintessence the perturbations drastically modify the properties and stability of the background evolution. It turns out taht in the quintessence model there is one and only one stable points. The behavior of this stable point leads either an exponentially increasing matter clustering, not detected in cosmological observations, or to a physically not interesting Laplacian instability. Hence, the quintessence cosmological models may be in a sever disadvantage as compared to the standard $\Lambda$CDM model. Some of these problems may be solvable in the dissipative quintessence scenario, which, for example, may limit the exponential increase of the matter clustering via the dissipation of the scalar field energy.

\section{ Discussions and final remarks}\label{sect5}

In the present paper we have considered the cosmological implications of a
dissipative scalar field, whose theoretical description can be obtained from
a variational principle, inspired by the case of the simple damped harmonic
oscillator. In performing such a generalization and extension of the scalar
field models we assume that dissipation may be a general property of
physical systems, and its presence should be unavoidable in any natural
process. It is interesting to note that at very low temperatures the
superfluid component of the liquid helium behaves as an irrotational ideal
fluid, flowing without friction \cite{D1,D2}. However, once a critical
velocity $v_c$ is reached, dissipation sets in, and the flow is not anymore
frictionless. In the standard physical interpretation of this process, it is
assumed that dissipation in the superfluid flow is due to the creation,
motion, and evolution of the superfluid quantized vortices in the liquid
\cite{D1,D2}. Dissipation can be generally attributed to the interaction of
the given physical system with an external (thermal, for example) bath, or
to the interaction with another physical system. The interaction between
dark energy and dark matter may provide a possible physical mechanism for
the presence of the dissipative effects of the two basic components of the
Universe.

Various forms of the dissipative Klein-Gordon equation have been
investigated, mostly from a mathematical point of view. The dissipative
Klein-Gordon equations are usually strongly non-linear partial differential
equations. An equation of the form,
\begin{equation}
\Box u +u=-g\left(\partial _t u\right)^2,
\end{equation}
where $g$ is a constant, and $\Box u=\partial _t^2-\partial _x^2$, was
investigated in \cite{M1}, where it was shown that the solution of the
non-linear equation has an additional logarithmic time decay in comparison
with the free evolution. The dissipative one-dimensional Klein-Gordon
equation,
\begin{equation}
u_t-u_{xx}+b(x)u_t+f(u)+h(\nabla u)=0,
\end{equation}
where $f$, $g$, $h$, $b$ are arbitrary functions, was studied in \cite{M2}.
A particular dissipative non-linear Klein-Gordon equation of the form,
\begin{equation}
u_{tt}-u_{xx}+\alpha u-\beta u^3=0,
\end{equation}
with $\alpha$ and $\beta $ constants, plays an important role in many fields
of physics, like in the study of the liquid helium, dislocations in
crystals, the Bloch wall motion, ferromagnetic materials, the unified theory
of elementary particles, Josephson array, charge density waves, the
propagation of magnetic flux on a Josephson line etc. (see \cite{M2} and
references therein). A non-linear dissipative Klein-Gordon equation, given
by,
\begin{equation}
u_{tt}-\Delta u+u+\gamma u_t=\left|u\right|^{p-1},
\end{equation}
was studied, from a mathematical point of view, in \cite{M3}. Hence, a large
number of dissipative Klein-Gordon type equations have been proposed, and
investigated in detail in both mathematical and physical literature.
However, most of these equations have been proposed on a phenomenological
basis, as mostly empirical models for the description of some physical
processes,

In dealing with the dissipation problem, in the present work we have
introduced a comprehensive description of the dissipative scalar fields,
based on a variational principle, which was inspired by the mathematics of
one of the simplest possible dissipative system, the damped harmonic
oscillator. Dissipative processes can also be described by variational
principles, even that these principles are not as commonly used as the
variational principles for conservative systems. However, the Lagrangians
for dissipative systems are almost as simple as those for conservative
systems, and, with the use of the Euler-Lagrange equations, they allow a
direct and systematic derivation of the equation of motion, and to obtain
the basic physical properties and characteristics of the dissipative systems.

In the present approach to the scalar field physics we have introduced theoretical models in which the ordinary Lagrangian of the field is multiplied by an arbitrary function of the coordinates, of the metric, and of the scalar field. The Euler-Lagrange equations straightforwardly lead to various dissipative formulations and extensions of the Klein-Gordon equation, whose forms depend now on the dissipation  exponent, and function. In a Riemannian geometry, the variational mathematical formalism allows to obtain the dissipative Klein-Gordon equations in an explicitly covariant form. The main goal of the present study was, besides introducing the theoretical formalism, to explore the implications of the dissipative scalar fields in cosmology. Scalar fields have been already extensively used as successful dark energy models, which can mimic/replace the cosmological constant, and thus provide powerful alternatives to the standard $\Lambda$CDM paradigm. In order to develop some cosmological applications, we have considered dissipative scalar field models leading to the generalized Klein-Gordon equation of the form $\ddot{\phi}+3H(1+Q)\dot{\phi}+V'(\phi)=0$, which was also considered previously in the framework of warm inflationary cosmological models, but without being derived from a variational principle \cite{W1, W2}. This dissipative Klein-Gordon equation can be derived from the standard Lagrangian $L_\phi=e^{3\int{H(t)Q(t)dt}}\rho_\phi$.

The variational principle allows not only the systematic introduction of the dissipation in scalar field models, but also obtaining the effective energy density and pressure that can be associated to the scalar field. The effective energy of the field can be obtained as the effective Hamiltonian derived in the standard way from the field Lagrangian. On the other hand, to obtained the effective pressure of the field we have imposed the cosmological conservation of the effective quantities. Generally, the Friedmann equations imply the conservation of the total matter-field content of the Universe. By imposing the independent conservation laws for matter and field we have neglected the possibility of any interaction between scalar field and cosmological matter, even that such a possibility cannot be ruled out a priori.

The generalized conservation equation, with the effects of the matter ignored, uniquely determines the effective pressure of the dissipative field in the form $p_\phi ^{(eff)}=(1+Q)\left(\dot{\phi}^2/2-V(\phi)\right)e^{3\int{H(t)Q(t)dt}}$. This effective field pressure, together with the effective density $\rho_\phi^{(eff}=\left(\dot{\phi}^2/2-V(\phi)\right)e^{3\int{H(t)Q(t)dt}}$ are the physical quantities that appear in the generalized Friedmann equations that describe the cosmological dynamics. From a mathematical pointy of view, the Friedmann equations become differential-integral equations, with the inclusion of the dissipative effects leading to a significant increase in the mathematical problem of the cosmological evolution. However, the cosmological problem is still solvable relatively straightforwardly for the considered dissipation exponent, since by means of simple mathematical transformations, one can reformulate the Friedmann-Klein-Gordon system in the redshift space as a first order differential dynamical system, whose solutions can be obtained easily numerically.
We have examined in detail several cosmological models, which were obtained for different choices of the dissipation function, and of the scalar field potential. From the point of view of the dissipation function, we have considered models with constant $Q$, and with $Q$ a particular function of the redshift. For the scalar field potential we have also adopted two forms only, $V(\phi)=0$, and $V(\phi)=m\phi^2/2$, respectively.

From a cosmological point of view, the most significant change in the modelling of dark energy comes from the expression of the effective pressure. First of all, successful cosmological models without the presence of the potential can be easily constructed, by assuming that the dissipation function satisfies the condition $1+Q<0$. With this choice the kinetic term of the pressure becomes positive in the second Friedmann equation, and an effective negative pressure of the form $p_\phi^{(eff)}=\left((1+Q)\dot{\phi}^2/2\right)e^{3\int{H(t)Q(t)dt}}$ can effectively trigger, and control, the accelerated expansion of the Universe, thus playing the role of the cosmological constant, and of the dark energy. On the other hand, for this model, the sign of the kinetic term in the effective energy of the field has the correct sign. Hence, no self-interacting potential is necessary for a dissipative scalar field to accelerate the Universe, the role of the potential being taken over by the dissipation function. On the other hand, while many fundamental physical models do exist for the scalar field potential, to the best knowledge of the present author, no theoretical models for the dissipation exponent have been considered in the framework of the fundamental theories of elementary particle physics.

It is important to point out that, even at low redshifts $z<2$ the predictions of the dissipative quintessence model do coincide with the predictions of the $\Lambda$CDM model, and with the observational data, some significant deviations may appear at higher redshifts $z>2.5$. For standard quintessence models, the deviations from the evolution of the $\Lambda$CDM are bounded to be below the 10\% level at 95\% confidence at redshifts below $z = 1.5$ \cite{new1}. It would be interesting to investigate if the inclusion of the dissipative processes of the scalar field could significantly change this bound. On the other hand, in the present models the dissipation function can be taken as an increasing function of the redshift (a decreasing function of time), and thus, at enough high redshifts, due to the presence of the function $e^{\Gamma}$ in the expressions of $\rho_\phi$ and $p_\phi$, in the early Universe the contributions of the scalar field energy density and pressure becomes negligible, and the Universe is matter dominated, with a decelerating evolution. Hence, generally, we expect that the dissipative quintessence evolution takes place in three phases. In the first phase, at low redshifts $z<2$, the model (almost) coincides with $\Lambda$CDM, and describes the present day accelerating evolution. At intermediate redshifts, in the (approximate) range  $2<z<5$, the dissipative quintessential cosmological expansion may differ, even significantly, from the $\Lambda$CDM evolution. However, at $z>5$, both models become matter dominated, and thus their large redshifts dynamics coincides again.
Hence,  the early matter dominated cosmological phase is recovered in a large redshift limit in the present model, due to the presence of the dissipation function in the expressions of the basic quantities describing the quintessence field, and without the necessity of introducing any supplementary assumptions, like, for example, a change of the potential, or specific initial conditions. Moreover, there are no restrictions on the scalar field potential, since the inclusion of a proper dissipation function in the scalar field equations would automatically recover the early matter dominated era.

In order to confront this theoretical model with the observations we have considered several simple models, obtained by assuming some simple forms for the dissipation function, and for the scalar field potential. All the considered cases have been compared with a (limited) set of observational data for the Hubble function, and with the predictions of the $\Lambda$CDM model. The generalized Friedmann equations have been solved numerically, with the initial conditions chosen for the scalar field, and its derivative so that the models comes as close as possible to the observations and to the $\Lambda$CDM model. I would like to point out that no fitting was used to obtain, and fix, the free parameters of the models, but the results have been obtained by the trial and search method. As a general conclusion of these investigations one can say that the dissipative scalar field model, in its various versions, can give a good description of the observational cosmological data, and succeeds in reproducing the predictions of the $\Lambda$CDM model. Of course, a detailed analysis of a larger number of cosmological data is necessary, before one could give a fair estimate of the potential of the dissipative scalar field cosmological models. And deep investigations into the origin, and physical mechanisms of dissipation, at both classical and quantum levels, are also necessary.

By taking into account the results of the present work, the dissipative scalar field cosmological models could become an attractive physical alternative to the standard $\Lambda$CDM model concerning the theoretical interpretation, and the explanation of the observational data. It
may also give a new vision, and a better comprehension of the complex, and unexpected, dynamical processes that take place in the Universe.

\section*{Acknowledgements}

The work of TH is supported by a grant of the Romanian Ministry of Education and Research, CNCS-UEFISCDI, project number PN-III-P4-ID-PCE-2020-2255 (PNCDI III).

\end{document}